
\input epsf
\input harvmac
\input amssym.def
\input amssym
\catcode`\:=11
\newcount\bm:counta \newcount\bm:countb 
\newcount\bm:countc \newcount\bm:countd
\newtoks\bm:tok
\newif\ifbm:delim
\def\thehex#1{\ifcase\the#1 0\or 1\or 2\or 3\or 4\or 5\or 6\or 7\or
8\or 9\or A \or B\or C\or D\or E\or F\fi}%
\def\test#1#2{\ifcat#1#2\message{True}\else\message{False}\fi}
\newtoks \bm:savedtoks  \bm:savedtoks{}%
\def\bm:empty{\relax}%
\let\bm:save=\bm:empty
\newif\ifbm:cdr
\def\bm:split#1#2\bm:empty{%
    \def\bm:car{#1}\def\bm:cdr{#2\relax}%
    \expandafter\ifx\expandafter\relax\bm:cdr\bm:cdrfalse
    \else\bm:cdrtrue
    \fi}%
\newif\ifbm:found
\def\bm:in#1\find:#2\this:{%
    \def\find:##1#2##2##3\find:{%
        \ifx\bm:in##2\bm:foundfalse
        \else\bm:foundtrue
        \fi}%
    \find:#1#2\bm:in\find:}%
\let\when=\relax \let\use=\relax
%
%
\newif\ifboldwarning
\def\bm:message#1{{\newlinechar=`^^J
\immediate\write16{\string\bold\space warning on line
                                             \the\inputlineno^^J#1^^J}}}%
%
\def\latex:adjust{\expandafter\ifx\the\textfont0\csname twlrm\endcsname
                                           \def\bm:scale{1200}%
                              \else\expandafter\ifx
                                 \the\textfont0\csname elvrm\endcsname
                                                 \def\bm:scale{1095}%
                                               \else\def\bm:scale{1000}%
                                               \fi
                              \fi}%
\latex:adjust
\newdimen\bm:sevensize \newdimen\bm:fivesize
\bm:sevensize=.007pt \bm:fivesize=.005pt
\bm:sevensize=\bm:scale\bm:sevensize
\bm:fivesize=\bm:scale\bm:fivesize
\font\tenbf=cmbx10 scaled \bm:scale
\font\sevenbf=cmbx7 at \the\bm:sevensize
\font\fivebf=cmbx5 at \the\bm:fivesize
\textfont\bffam=\tenbf
\scriptfont\bffam=\sevenbf
\scriptscriptfont\bffam=\fivebf
:bit=cmbxti10 scaled \bm:scale
:bit=cmbxti10 at \the\bm:sevensize
:bit=cmbxti10 at \the\bm:fivesize
:scale
 at \the\bm:sevensize
 at \the\bm:fivesize
:bsf=cmssbx10 scaled \bm:scale
:bsf=cmssbx10 at \the\bm:sevensize
:bsf=cmssbx10 at \the\bm:fivesize
:bsl=cmbxsl10 scaled \bm:scale
:bsl=cmbxsl10 at \the\bm:sevensize
:bsl=cmbxsl10 at \the\bm:fivesize
:bmit=cmmib10 scaled \bm:scale
:bmit=cmmib7 at \the\bm:sevensize
:bmit=cmmib5 at \the\bm:fivesize
\newfam\bm:bmitfam
\textfont\bm:bmitfam=\tenbm:bmit
\scriptfont\bm:bmitfam=\sevenbm:bmit
\scriptscriptfont\bm:bmitfam=\fivebm:bmit
:bsy=cmbsy10 scaled \bm:scale
:bsy=cmbsy7 at \the\bm:sevensize
:bsy=cmbsy5 at \the\bm:fivesize
\newfam\bm:bsyfam
\textfont\bm:bsyfam=\tenbm:bsy
\scriptfont\bm:bsyfam=\sevenbm:bsy
\scriptscriptfont\bm:bsyfam=\fivebm:bsy
\newtoks\alphatok \alphatok{?}%
\expandafter\def\expandafter\new:fam\expandafter{\newfam}%
\def\set:fam#1{%
    \expandafter\ifx\csname #1fam\endcsname\relax
                    \expandafter\new:fam\csname #1fam\endcsname
                    \edef\alphafam{\csname #1fam\endcsname}%
                    \edef\set:fonts{%
                         \global\textfont\alphafam=\csname ten#1\endcsname
                         \global\scriptfont\alphafam=\csname
                                                seven#1\endcsname 
                         \global\scriptscriptfont\alphafam=\csname
                                                        five#1\endcsname}%
                    \set:fonts
                \fi
}%
\def\declare:alpha#1{%
    \set:fam{#1}%
    \expandafter\edef\csname math#1\endcsname{{%
                     \noexpand\if?\noexpand\the\noexpand\alphatok
                                  \global\noexpand\bm:savedtoks
                            {\noexpand\csname math:#1\noexpand\endcsname}%
                                  \noexpand\aftergroup\noexpand\bm:getarg
                     \noexpand\else\errmessage{You're already inside a
                                   \noexpand\expandafter\noexpand\string
                                   \noexpand\the\alphatok{..} - don't
                                   even think about it!}%
                     \noexpand\fi}}%
    \expandafter\def\csname math:#1\endcsname##1{{%
                     \alphatok\expandafter{\csname math#1\endcsname}%
                     \fam\csname #1fam\endcsname
                     \let\boldletter=\relax ##1}}}%
%
\chardef\rmfam=0
\chardef\mitfam=1
\chardef\calfam=2
\declare:alpha{rm}%
\declare:alpha{it}%
\declare:alpha{sl}%
\declare:alpha{tt}%
\declare:alpha{bf}%
\declare:alpha{mit}%
\declare:alpha{cal}%
\declare:alpha{sf}
\declare:alpha{bm:bmit}
\declare:alpha{bm:bsy}
\def\default{\fam=-1 \def\boldletter##1{{\bf ##1}}}%
\let\mathbm=\mathbm:bmit :bsy
\let\mathbm:bcal=\mathbm:bsy
\def\bold#1{{
    \let\bm:currentsymbol=\relax
    \bm:split#1\bm:empty
    \ifbm:cdr\toks0{}\loop\toks0\expandafter\expandafter\expandafter{\expandafter\the\expandafter\toks0\expandafter\noexpand\expandafter\bold\expandafter{\bm:car}}\expandafter\bm:split\bm:cdr\bm:empty
                      \ifbm:cdr
                      \repeat
              \toks2\expandafter{\bm:car}%
              \xdef\bm:out{\the\toks0
                           \noexpand\bold\expandafter{\the\toks2}}%
              \aftergroup\bm:out 
    \else\ifx\bold#1\aftergroup\bold
         \else\ifmmode\bm:select{#1}%
                      \ifx\bm:save\bm:empty
                          \xdef\bm:out{\global\bm:savedtoks{}%
                             \the\bm:savedtoks\noexpand\bm:currentsymbol}%
                          \aftergroup\bm:out
                      \else\global\bm:savedtoks\expandafter\expandafter
          \expandafter{\expandafter\the\expandafter\bm:savedtoks\bm:save}%
                            \ifx\bm:currentsymbol\relax\aftergroup\bold
                            \else\let\test=F%
                                 \edef\argtest{\noexpand\bm:in
                                      \meaning\bm:currentsymbol
                                      \noexpand\find:
                                      \string\mathaccent
                                      \noexpand\this:}%
                                 \argtest
                                 \ifbm:found \let\test=T%
                                 \fi
                                 \edef\argtest{\noexpand\bm:in
                                      \meaning\bm:currentsymbol
                                      \noexpand\find:
                                      \string\radical
                                      \noexpand\this:}%
                                 \argtest
                                 \ifbm:found \let\test=T%
                                 \fi
                                 \if T\test\aftergroup\bm:getarg
                                 \else\aftergroup\bm:dumpchars
                                 \fi
                            \fi
                      \fi
              \else\errmessage{\string\bold\space should be used in
                               math mode only}%
              \fi
          \fi
    \fi
    }}%
\def\bm:getarg#1{{%
    \ifx#1\bold\aftergroup\bold
    \else\toks0{#1}\xdef\bm:out{\global\bm:savedtoks{}%
                        \the\bm:savedtoks{\the\toks0}}%
         \aftergroup\bm:out
    \fi}}%
\def\bm:dumpchars{{%
    \xdef\bm:out{\global\bm:savedtoks{}\the\bm:savedtoks}%
    \aftergroup\bm:out}}%
\def\bm:select#1{%
    \expandafter\bm:in\boldspecials\find:#1\this:
    \ifbm:found \def\when##1\use##2;{%
                    \ifx##1#1\xdef\bm:currentsymbol{\noexpand##2}%
                    \else\ifx##2#1\xdef\bm:currentsymbol{\noexpand##2}\fi
                    \fi}%
                \boldspecials
    \else\if\noexpand#1\relax                    
            \let\test=F%
            \edef\chartest{\noexpand\bm:in\meaning#1\noexpand\find:
                                         \string\mathchar\noexpand\this:}%
            \chartest
            \ifbm:found\let\test=T%
            \fi
            \edef\chartest{\noexpand\bm:in\meaning#1\noexpand\find:
                                           \string\char\noexpand\this:}%
            \chartest
            \ifbm:found \let\test=T%
            \fi                                  
            \if T\test 
          \expandafter\ifx\csname bold\string#1\endcsname\relax 
                          \edef\bm:process{\noexpand\defboldsymbol
                                           {\noexpand#1}%
                                           {\noexpand\mathchar}%
                                           {\the#1}{-1}}%
                          \bm:process
                       \fi \global\expandafter\let\expandafter
                           \bm:currentsymbol\expandafter=%
                           \csname bold\string#1\endcsname
            \else 
          \expandafter\ifx\csname bold\string#1\endcsname\relax 
                          \ifboldwarning\bm:message{Skipping \string#1.}%
                          \fi
                                        \def\bm:save{#1}%
                      \else\toks4\expandafter{%
                           \csname bold\string#1\endcsname}%
                           \edef\bm:save{\the\toks4}%
                           \global\expandafter\let\expandafter
                           \bm:currentsymbol\expandafter=%
                           \the\toks4
                      \fi
            \fi
         \else                                   
              \ifcat\noexpand#1A\gdef\bm:currentsymbol{\boldletter{#1}}%
              \else\ifcat\noexpand#1>            
             \expandafter\ifx\csname bold\string#1\endcsname\relax
                             \edef\bm:process{\noexpand\defboldsymbol
                                              {\noexpand#1}%
                                              {\ifnum\the\delcode`#1>-1
                                                 \noexpand\delimiter
                                               \else\noexpand\mathchar
                                               \fi}%
                                              {\the\mathcode`#1}%
                                              {\the\delcode`#1}}%
                             \bm:process         
                          \fi\global\expandafter\let
                             \expandafter\bm:currentsymbol
                             \expandafter=\csname 
                             bold\string#1\endcsname
                   \else\ifboldwarning\bm:message{Skipping \string#1.}%
                        \fi
                                      \def\bm:save{#1}%
                   \fi
              \fi
         \fi
    \fi}%
\def\defboldsymbol#1#2#3#4{%
    \bm:tok={#1}%
 \expandafter\ifx\csname bold\string#1\endcsname\relax
             \else\ifboldwarning\bm:message{Redefining
                                                \string\bold\string#1.}%
                  \fi
             \fi
    \bm:counta=#3
    \bm:countd=#4
    \ifnum\the\bm:counta="8000
           \expandafter\xdef\csname bold\string#1\endcsname{#1}%
    \else\ifx#2\delimiter \bm:delimtrue
         \else\ifx#2\radical \bm:delimtrue
              \else \bm:delimfalse
              \fi
         \fi
         \bm:countc=\bm:counta
         \divide\bm:countc by "1000             
         \advance\bm:counta by -\expandafter"\thehex\bm:countc 000
         \ifbm:delim\ifnum\the\bm:countd>-1     
                          \begingroup \bm:counta=\bm:countd
                          \divide\bm:counta by "1000
                              \begingroup
                              \multiply\bm:counta by "1000
                              \global\advance\bm:countd by
                                                      -\the\bm:counta
                              \endgroup
                          \bold:mathrecode
                          \multiply\bm:counta by "1000
                              \begingroup
                              \bm:counta=\bm:countd
                              \bold:mathrecode
                              \global\bm:countd=\bm:counta
                              \endgroup
                          \global\advance\bm:countd by \the\bm:counta
                          \endgroup
                      \else\ifboldwarning\bm:message{\the\bm:tok\space
                        is not a \string#2.^^JDoing the obvious thing..}%
                           \fi
                      \bold:mathrecode
                      \bm:countd=\bm:counta
                      \multiply\bm:counta by "1000
                      \advance\bm:countd by \the\bm:counta
                      \fi
                      \advance\bm:countd by \expandafter"\thehex
                                                        \bm:countc000000
                      \expandafter\xdef\csname bold\string#1\endcsname
                               {#2\the\bm:countd}%
          \else\bold:mathrecode
               \advance\bm:counta by \expandafter"\thehex\bm:countc 000
               \ifx#2\mathchar
                   \global\expandafter\mathchardef\csname
                                bold\string#1\endcsname=\the\bm:counta
               \else\expandafter\xdef\csname bold\string#1\endcsname
                               {#2\the\bm:counta}%
               \fi
          \fi
    \fi
}%
\def\bold:mathrecode{
      \bm:countb=\bm:counta
      \divide\bm:countb by "100
      \advance\bm:counta by -\expandafter"\thehex\bm:countb 00
      \ifcase\the\bm:countb
              \bm:countb=\the\bffam
      \or     \bm:countb=\the\bm:bmitfam
      \or     \bm:countb=\the\bm:bsyfam
      \else   \ifboldwarning\ifbm:delim\bm:message{Lack of bold
                                        extension fonts means
                                        \string\bold\the\bm:tok\space may
                                        not be bold.}%
                             \else\bm:message{Sorry, there just aren't the
                                   fonts for \string\bold\the\bm:tok.}%
                             \fi
              \fi 
      \fi
\advance\bm:counta by \expandafter"\thehex\bm:countb 00
                    }%
\def\DeclareBoldMacro#1#2#3{
    \bm:counta=#3 \bm:countd=\bm:counta
    \ifnum\the\bm:counta>"7FFF 
          \divide\bm:counta by "1000
    \else
          \multiply\bm:countd by "1000
    \fi
    \bm:countc=\bm:counta
    \divide\bm:countc by "1000
    \advance\bm:countd by -"\thehex\bm:countc 000000
    \edef\bm:process{\noexpand\defboldsymbol{\noexpand#1}{\noexpand#2}%
                     {\the\bm:counta}{\the\bm:countd}}%
    \bm:process}%
%
%
\DeclareBoldMacro{\{}{\delimiter}{"4266308}
\DeclareBoldMacro{\}}{\delimiter}{"5267309}
\DeclareBoldMacro{\langle}{\delimiter}{"426830A}
\DeclareBoldMacro{\rangle}{\delimiter}{"526930B}
\edef\FixLessThanGreaterThan{\noexpand\DeclareBoldMacro{<}%
{\noexpand\mathchar}{\the\mathcode`\<}%
\noexpand\DeclareBoldMacro{>}{\noexpand\mathchar}{\the\mathcode`\>}}%
\FixLessThanGreaterThan
\DeclareBoldMacro{\sqrt}{\radical}{"270370}
\default
\def\boldspecials{\when\mathmit\use\mathbm:bmit;\when\mathcal\use\mathbm:bcal;\when\mathrm\use\mathbf;\when\mathsf\use\mathbm:bsf;\when\mathit\use\mathbm:bit;\when\mathtt\use\mathtt;\when\mathsl\use\mathbm:bsl;\when\default\use\default;}%
\boldwarningtrue
\catcode`\:=12

\magnification\magstep1
\baselineskip 12pt
\hyphenation{Wissen-schaf-ten}
\hyphenation{Min-kow-ski}

\newdimen\itemindent \itemindent=32pt
\def\textindent#1{\parindent=\itemindent\let\par=\resetpar%
\indent\llap{#1\enspace}\ignorespaces}

\let\oldpar=\par
\def\resetpar{\oldpar\parindent=20pt\let\par=\oldpar}

\font\ninerm=cmr9 \font\ninesy=cmsy9
\font\eightrm=cmr8 \font\sixrm=cmr6
\font\eighti=cmmi8 \font\sixi=cmmi6
\font\eightsy=cmsy8 \font\sixsy=cmsy6
\font\eightbf=cmbx8 \font\sixbf=cmbx6
\font\eightit=cmti8
\def\eightpoint{\def\rm{\fam0\eightrm}
  \textfont0=\eightrm \scriptfont0=\sixrm \scriptscriptfont0=\fiverm
  \textfont1=\eighti  \scriptfont1=\sixi  \scriptscriptfont1=\fivei
  \textfont2=\eightsy \scriptfont2=\sixsy \scriptscriptfont2=\fivesy
  \textfont3=\tenex   \scriptfont3=\tenex \scriptscriptfont3=\tenex
  \textfont\itfam=\eightit  \def\it{\fam\itfam\eightit}%
  \textfont\bffam=\eightbf  \scriptfont\bffam=\sixbf
  \scriptscriptfont\bffam=\fivebf  \def\bf{\fam\bffam\eightbf}%
  \normalbaselineskip=9pt
  \setbox\strutbox=\hbox{\vrule height7pt depth2pt width0pt}%
  \let\big=\eightbig  \normalbaselines\rm}
\catcode`@=11 %
\def\eightbig#1{{\hbox{$\textfont0=\ninerm\textfont2=\ninesy
  \left#1\vbox to6.5pt{}\right.\n@space$}}}
\def\vfootnote#1{\insert\footins\bgroup\eightpoint
  \interlinepenalty=\interfootnotelinepenalty
  \splittopskip=\ht\strutbox %
  \splitmaxdepth=\dp\strutbox %
  \leftskip=0pt \rightskip=0pt \spaceskip=0pt \xspaceskip=0pt
  \textindent{#1}\footstrut\futurelet\next\fo@t}
\catcode`@=12 %

\font \bigbf=cmbx10 scaled \magstep1
\def \bnab{{\bold \nabla}}
\def \de{\delta}
\def \si{\sigma}

\def \nab{\nabla}
\def \pr{\partial}
\def \bx{{\bf x}}

\def \d{{\rm d}}
\def \bA{{\bf A}}
\def \rO{{\rm O}}

\def \half{{\textstyle {1 \over 2}}}
\def \quar{{\textstyle {1 \over 4}}}

\def \ts{ \textstyle}

\def \tt{{\tilde t}}
\def \tx{{\tilde x}}
\def \ba{{\bf a}}
\def \btx{{\tilde{\bf x}}}
\def \A{{\cal A}}
\def \B{{\cal B}}

\def \F{{\cal F}}
\def \G{{\cal G}}
\def \I{{\cal I}}
\def \J{{\cal J}}
\def \K{{\cal K}}
\def \L{{\cal L}}

\def \P{{\cal P}}
\def \Q{{\cal Q}}
\def \O{{\cal O}}
\def \R{{\cal R}}
\def \T{{\cal T}}
\def \U{{\cal U}}
\def \V{{\cal V}}
\def \W{{\cal W}}
\def \X{{\cal X}}
\def \Y{{\cal Y}}
\def \Z{{\cal Z}}

\def \ep{\epsilon}

\lref\Bate{H. Bateman, Proc. London Math. Soc. 8 (1910) 223.}
\lref\Cunn{E. Cunningham, Proc. London Math. Soc. 8 (1910) 77.}
\lref\Page{L. Page and N.I. Adams, Phys. Rev. 49 (1936) 466.}
\lref\acc{T. Fulton, F. Rohrlich and L. Witten, Nuovo Cimento 26 (1962) 652\semi
F. Rohrlich, Ann. Phys. (N.Y.) 22 (1963) 169.}
\lref\Born{M. Born, Ann. Physik 30 (1909) 1.}
\lref\Fron{B. Binegar, C. Fronsdal and W. Heidenreich, 
J. Math. Phys. 24 (1983) 2826.}
\lref\Dirac{P.A.M. Dirac, Annals of Math. 37 (1936) 823; {\it in} ``The
Collected Works of P.A.M. Dirac 1924-1948'', edited by R.H. Dalitz, CUP 
(Cambridge) 1995.}
\lref\Bondi{H. Bondi and T. Gold, Proc. Roy. Soc. A229 (1955) 416.}
\lref\consv{D.G. Boulware, L.S. Brown and R.D. Peccei,
Phys. Rev. D2 (1970) 293.}
\lref\conf{H.A. Kastrup, Phys. Rev. 150 (1964) 1183\semi
G. Mack and A. Salam, Ann. Phys. (N.Y.) 53 (1969), 174\semi
S. Ferrara, R. Gatto and A.F. Grillo, ``Conformal Algebra in Space-time and
Operator Product Expansion", (Springer Tracts in Modern Physics, vol. 67)
Springer (Heidelberg) 1973\semi
D.H. Mayer, J. Math. Phys. 16 (1975) 884\semi
F. Bayen, M. Flato, C. Fronsdal and A. Haidari, Phys. Rev. D32 (1985) 2673\semi
P. Furlan, V.B. Petkova, G.M. Sotkov and I.T. Todorov, Revista del Nuovo
Cimento 8 (1985) 1\semi
V.B. Petkova, G.M. Sotkov and I.T. Todorov, Comm. Math. Phys. 97 (1985) 227.}
\lref\Cast{L. Castell, Nucl. Phys. B13 (1969) 231.}
\lref\acc{G.A. Schott, ``Electromagnetic Radiation'', CUP (Cambridge) 1912\semi
W. Pauli, Relativit\"atstheorie
{\it in} ``Enzyklop\"adie der mathematischen Wissenschaften'', sect. V, 
vol. 19, part 2, p. 539, B.G. Teubner (Leipzig) 1921; translated in W. Pauli, 
``Theory of Relativity'',
Pergamon Press (London) 1958, pp. 92-94\semi
S.R. Milner, Phil. Mag. 41 (1921) 405\semi
D.L. Drukey, Phys. Rev. 76 (1949) 543\semi
T. Fulton and F. Rohrlich, Ann. Phys. (N.Y.) 9 (1960) 499\semi
J. Cohn, Am. J. Phys. 46 (1978) 225\semi
R.E. Peierls, ``Surprises in Theoretical Physics'', pp. 160-166, Princeton
University Press (Princeton) 1979\semi
D.G. Boulware, Ann. Phys. (N.Y.) 124 (1980) 169\semi
H. Bondi,  Proc. Roy. Soc. A376 (1981) 493\semi
T. Nagatsuka and S. Takagi, Ann. Phys. (N.Y.) 242 (1995) 292.}
\lref\Hill{E.L. Hill, Phys. Rev. 72 (1947) 143.}
\lref\Adler{S.L. Adler, Phys. Rev. D6 (1972) 3445, E D7 (1973) 3821; D8 (1973)
2400.}
\lref\Shore{I.T. Drummond and G.M. Shore, Ann. Phys. (N.Y.) 117 (1979) 
89, 121\semi
G.M. Shore, Phys. Rev. D21 (1980) 2226; Ann. Phys. (N.Y.) 122 (1980) 321.}
\lref\Baker{M. Baker and K. Johnson, Physica 96A (1979) 120\semi
J. Erhlich and D.Z. Freedman, preprint MIT-CTP-2588, hep-th/9611133.}
\lref\Witten{P.C. Argyres, M.R. Plesser, N. Seiberg and E. Witten, Nucl. Phys.
{B461} (1996) 71, hep-th/9511154.}
\lref\confb{G.M. Sotkov and D.T. Stoyanov, J. Phys. A 13 (1980) 2807; 16
(1983) 2817\semi
E.S. Fradkin, A.A. Kozhevnikov, M.Ya. Palchik and A.A. Pomeransky,
Comm. Math. Phys. 91 (1983) 529\semi
R.P. Zaikov, Theor. and Math. Phys. 65 (1985) 1012, 67 (1986) 368; Lett.
in Math. Phys. (1986) 189\semi
S. Ichinose, Lett. in Math. Phys. 11 (1986) 113; Nucl. Phys.
272 (1986) 727\semi
A.D. Haidari, J. Math. Phys. 27 (1986) 2409.}
\lref\confc{H.J. Schnitzer, Phys. Rev. D8 (1973) 385\semi
N. Christ, Phys. Rev. D9 (1974) 946\semi
M.P. Fry, Nuovo Cimento 31A (1976) 129; Nucl. Phys. B107 (1976) 535,
B121 (1977) 343.}

\parskip 5pt
{\nopagenumbers
\rightline{hep-th/9701064}
\vskip 2truecm
\centerline {\bigbf Conformal Invariance and Electrodynamics:}
\vskip 5pt
\centerline {\bigbf Applications and General Formalism}
\vskip 2.0 true cm
\centerline {C. Codirla and H. Osborn\footnote{*}{email: 
cc10011@cus.cam.ac.uk and ho@damtp.cam.ac.uk}}
\vskip 10pt
\centerline {Trinity College}
\centerline {Cambridge, CB2 1TQ}
\centerline {England}
\vskip 2.0 true cm
{\eightpoint
\parindent 1.5cm{
{\narrower\smallskip\parindent 0pt
The role of the conformal group in electrodynamics in four space-time
dimensions is re-examined. As a pedagogic example we use the application of 
conformal transformations
to find the electromagnetic field for a charged particle moving with a
constant relativistic acceleration from the Coulomb electric field for the
particle at rest. We also re-consider the reformulation of Maxwell's
equations on the projective cone, which is isomorphic to a conformal
compactification on Minkowski space, so that conformal transformations,
belonging to the group $O(4,2)$, are realised linearly. The resulting
equations are different from those postulated previously and respect
additional gauge invariances which play an essential role in ensuring
consistency with conventional electrodynamics on Minkowski space. The solution
on the projective cone corresponding to a constantly accelerating charged
particle is discussed.

\narrower}}}
\vfill\eject}
\pageno=1
\noindent{\bigbf 1 Introduction}
\medskip
The invariance of Maxwell's equations under Lorentz transformations is the
cornerstone of the theory of relativity as expounded in the epoch making
paper of Einstein in 1905. In 1909 Cunningham \Cunn\ and Bateman \Bate\
showed that 
Maxwell's equations were also invariant under the larger conformal group. This
invariance does not extend to theories containing any mass scale
and is violated in quantum field theories, even when true classically,
except under very special circumstances so this symmetry has not been played
a significant role in mainstream theoretical
physics. Nevertheless a particular feature of the conformal group is that it
extends the usual Lorentz group
in allowing transformations to frames undergoing constant acceleration. In
the next section show how the conformal group can be used to obtain expressions
for the electric and magnetic fields, known also since 1909, corresponding to
a charged particle which undergoes constant acceleration or hyperbolic motion. 
The fields obtained by conformal transformation are non zero everywhere
for all time and are of course solutions of Maxwell's equations. They
are related to, but not identical with, the standard retarded, or advanced, 
solutions since these are zero on half of space-time. We also
describe briefly the derivation of these retarded, advanced solutions together
with the additional terms which are necessary to ensure they satisfy
Maxwell's equations despite the presence of discontinuities. In section 3 we
then recapitulate the well known relation of conformal transformations to
linear transformations on a projective cone and in section 4 we discuss
how electrodynamics may be reformulated equivalently in a explicitly
conformal invariant fashion in terms of fields on the projective cone.
Although the starting point is essentially identical with that originated  by
Dirac \Dirac\ and developed subsequently by various authors \refs{\conf,\Fron}
the equations derived
here are rather different. They are well defined on the projective cone and
maintain manifest gauge invariance, including the additional gauge 
transformations which are essential to ensure an exact equivalence between
fields on Minkowski space and the projective cone.
We then show in section 5 how the solution corresponding to an accelerating 
charged particle
may be simply expressed in terms of fields on the projective cone satisfying
corresponding versions of Maxwell's equations. In section 6 some more
general aspects of our results are discussed. Various detailed calculations
are relegated to an appendix.
\bigskip
\noindent{\bigbf 2 Conformal Transformations and Accelerating Charged Particles}
\medskip
We first define the conformal group as those transformations which leave the
relativistic line element invariant up to a factor so that if $\tx \to x$
\footnote{${}^1$}{As usual we set $c=1$ and adopt a metric with
$g_{\mu\nu}={\rm diag.}(-1,1,1,1)$.}
$$
g_{\mu\nu}\d \tx^\mu \d \tx^\nu = \Omega(\tx)^2 \, 
g_{\mu\nu}\d x^\mu \d x^\nu \, .
\eqno (2.1) $$
Obviously this includes translations, spatial rotations and Lorentz
transformations, for which $\Omega =1$, and scale transformations given by
$x = \rho \tx$. There are also inversions, which may be taken to have
the form
$$
x^\mu = {\tx^\mu \over \tx^2} \, , \qquad \Omega(\tx) = \tx^2 \, .
\eqno (2.2) $$
we may then define special conformal transformations by combining an
inversion with a translation and then another inversion,
$$  \eqalign{
\tx^\mu \to {\tx^\mu \over \tx^2}  \to {}& {\tx^\mu \over \tx^2} + b^\mu
\to \Big ( {\tx^\mu \over \tx^2} + b^\mu \Big ) \bigg / \Big (
{\tx \over \tx^2} + b \Big )^{\! 2} 
= {\tx^\mu + \tx^2 b^\mu \over \Omega(\tx)} = x^\mu \, , \cr
& \Omega(\tx) = 1 + 2b{\cdot \tx} + b^2 \tx^2 = (1-2b{\cdot x}+b^2x^2)^{-1}
\, . \cr}
\eqno(2.3) $$
Transformations such as (2.2) or (2.3) do not in general preserve the time
ordering of even time-like separated points. As shown in section 3
the action of conformal transformations
is transitive on a compactification of Minkowski space-time which has the
topology of $S^1 \times S^3/Z_2$, with $S^n$ the $n$-dimensional sphere and
$Z_2$ denoting the group formed by reflections and the identity.

To understand the significance of (2.3) let us set $b^\mu = ( 0, -\half \ba)$
and with $x^\mu = (t,\bx)$ we obtain
$$
\bx = {1\over \Omega(\tx)} \big ( \btx + \half \ba (\tt^2 -\btx^2 ) \big )\, ,
\quad t = {1\over \Omega(\tx)} \tt \, , \quad 
\Omega(\tx) = 1 - \ba{\cdot \btx} + \quar a^2 (\btx^2 - \tt^2 ) \, .
\eqno (2.4) $$
The spatial origin $\btx = {\bf 0}$ then transforms to
$$
\bx_o = {\half \ba \, \tt^2 \over 1- \quar a^2 \tt^2} \, , \qquad
t_o = {\tt  \over 1- \quar a^2 \tt^2} \, ,
\eqno (2.5) $$
so that $\bx_o$ and $t_o$ lie on a hyperbolic curve specified by the equations
$$
\bx_o = -\half \ba (\bx_o{}^{\! 2} - t_o{}^{\! 2}) \, \qquad  \hbox{or} \qquad
\bx_o \parallel \ba \, , \quad
(1+  \ba{\cdot \bx}_o )^2 - a^2 t_o {}^{\! 2} = 1 \, .
\eqno (2.6) $$
For $|\tt | < 2/a$ (2.5) represents the coordinates of a point moving with
constant acceleration $\ba$ passing through the origin $\bx_o = {\bf 0}$ at
$t_o=0$.\footnote{${}^2$}{The conformal transformation (2.4) is of course not
the same as that which gives rise to a static homogeneous gravitational field
\acc.}

The transformation of the electromagnetic fields which ensures that
Maxwell's equations are invariant is simply given by
$$
F_{\mu\nu}(x) \d x^\mu \d x^\nu = {\widetilde F}_{\mu\nu}(\tx) 
\d \tx^\mu \d \tx^\nu \, ,
\eqno (2.7) $$
or alternatively using (2.1)
$$
F^{\mu\nu}(x) = \Omega(\tx)^4 \, {\pr x^\mu \over \pr \tx^\si}
{\pr x^\nu \over \pr \tx^\rho} {\widetilde F}^{\si\rho}(\tx) \, .
\eqno (2.8) $$
Since $\d^4 x = \Omega(\tx)^{-4} \d^4 \tx$ it is easy to see that (2.7) and
(2.8) ensure invariance of the usual relativistic action 
$$
- {1\over 4}\int \! \d^4 x \, F^{\mu\nu}F_{\mu\nu}\, ,
\eqno (2.9) $$
whose variation leads to Maxwell's equations expressed in relativistically
covariant form. For the transformation (2.4) we may find
$$ \eqalign {
{\pr x^\mu \over \pr \tx^\si} = {}& {1\over \Omega(\tx)^2}
\pmatrix{ \Omega(\tx) + \half a^2 \tt^2 & \tt \, V_j \cr
\tt \, U^i & \Omega(\tx) \Big ( \de^i{}_{\! j} - 2 {a^i a_j \over a^2}\Big )
+ {2\over a^2} U^i V_j \cr} \, , \cr
& {\bf U} = (1-\ba{\cdot \btx}) \ba + \half a^2  \btx \, , \quad
{\bf V} = \ba - \half a^2  \btx \, . \cr}
\eqno (2.10) $$
Using these results we may find the transformation properties of electric and
magnetic fields, $E_i= F^{0i}$ and $B_i=\half \ep_{ijk}F^{jk}$, between
frames undergoing constant acceleration. For convenience we set 
$\widetilde{\bf B}={\bf 0}$ and then from (2.8,9) we get\footnote{${}^3$}{
Alternatively the line $\btx = {\bf 0}$ is mapped to the hyperbola (2.6)
just by the inversion $\bx = (\btx - {1\over 2} \ba)/\Omega$, $t = \tt/\Omega$
with $\Omega = (\btx - {1\over 2} \ba)^2 - \tt^2 = (\bx^2 - t^2)^{-1}$.
In this case the fields transform more simply as ${\bf B}(x) = 2\Omega^3
t \, \bx \times \widetilde{\bf E}(\tx)$, ${\bf E}(x) = \Omega^3 \big \{
(\bx^2 + t^2) \widetilde{\bf E}(\tx) - 2 \bx \, \bx{\cdot 
\widetilde{\bf E}}(\tx) \big \} $. For $\ba={\bf 0}$ this result was found in
\Cunn\ and later obtained, without making the connection to conformal
transformations, in ref. \Page.}
$$ \eqalign{
{\bf B}(x) ={}& \tt \, \Omega(\tx)  \big \{ (1- \ba{\cdot \btx})\,
\ba \times \widetilde{\bf E}(\tx)  +\half a^2 \, \btx \times 
\widetilde{\bf E}(\tx)
- \btx \times \ba \, \ba{\cdot \widetilde{\bf E}}(\tx) \big \} \, , \cr
{\bf E}(x) ={}& \Omega(\tx)^2 \widetilde{\bf E}(\tx) + \half \tt^2 \, 
\Omega(\tx)
( a^2 \widetilde{\bf E}(\tx) - \ba\, \ba{\cdot \widetilde{\bf E}}(\tx) ) \cr
& - \Omega(\tx) \big \{ (1- \ba{\cdot \btx})\, \ba \, 
\btx {\cdot \widetilde{\bf E}}(\tx) 
- \btx \, \ba{\cdot \widetilde{\bf E}}(\tx)
+ \half a^2 \btx\, \btx {\cdot \widetilde{\bf E}}(\tx) + \half \btx^2 \ba\,
\ba{\cdot \widetilde{\bf E}}(\tx) \big \} \, . \cr}
\eqno (2.11) $$
As a consistency check from (2.11) we may verify that ${\bf B}{\cdot {\bf E}}
=0$ and also ${\bf E}^2 - {\bf B}^2 = \Omega^4 {\widetilde{\bf E}}^2$.

The general result in (2.11) is rather complicated but as a simple application
we consider starting from the static Coulomb field for a point charge at
$\btx = {\bf 0}$,
$$ \widetilde{\bf E}(\tx) = {e\over 4\pi}\, {\btx \over |\btx|^3} \, .
\eqno (2.12) $$
Applying (2.11) and also (2.5) to eliminate $\tt, \btx$ gives
$$ \eqalign{
{\bf B}(x) ={}& {e\over 4\pi}\,
{t\, \ba \times \bx\over | \bx + \half \ba (\bx^2 - t^2)|^3}\, ,\cr
{\bf E}(x) ={}& {e\over 4\pi}\,
{ \bx (1+ \ba {\cdot \bx}) - \half \ba (\bx^2 + t^2 )
\over | \bx + \half \ba (\bx^2 - t^2)|^3 } \, . \cr}
\eqno (2.13) $$
In obtaining these results we have discarded a factor $\ep(\Omega)$, where
$\ep(x) = \pm 1$ for ${x\gtrless 0}$, or equivalently we have tacitly 
assumed that $\Omega>0$ when, according to (2.4), the transformation preserves
the time direction. We discuss this further in section 6. For the moment,
with transformation in (2.4), we may note that
$\Omega^{-1} = 0$ on the light cone through $\bx = - 2\ba/a^2, \, t=0$ and
$\Omega<0$ in the interior of this light cone.\footnote{${}^4$}{If the
transformation in the previous footnote is applied to (2.12) then the result
(2.13) is again obtained with a factor $-\ep(\Omega)$, but in this case
$\Omega<0$ in the interior of the light cone through $\bx={\bf 0}, \, t=0$.}
The fields in (2.13) are singular if $\bx = - \half \ba (\bx^2-t^2)$ 
so that $\bx,t$ then satisfy the equation of the hyperbola
given by (2.6). This equation has two branches although $\Omega>0$ only on the
curve passing through the space-time origin corresponding to $|\tt | < 2/a$.
The solution (2.13) satisfies the reflection symmetry
$$
{\bf B}(\bx,t) = - {\bf B}(-\bx-2\ba/a^2,t) \, , \quad
{\bf E}(\bx,t) =  {\bf E}(-\bx-2\ba/a^2,t) \, ,
\eqno (2.14) $$
and as a consequence it is easy
to see that (2.13) corresponds to a particle of charge $e$ moving on the 
hyperbola passing through the origin and charge $-e$ on the other branch 
where $\bx=-2\ba/a^2$ at $t=0$. For $|\bx|\sim t
\to \infty$ the electric, magnetic fields given by (2.13) reduce to standard
results for the electromagnetic waves radiated by an accelerating charged
particle. The solution (2.13) is usually ascribed to Born \Born\
and has been the subject of continued discussion \refs{\acc,\Bondi}
concerning its physical significance. A similar application of conformal
transformations was seemingly undertaken long ago by Hill.\footnote{${}^5$}
{See footnote 25 in ref. ({\Hill}).}

We may also consider applying a conformal transformation to the charge-current
density for which
$$
J^\mu(x) = \Omega(\tx)^4 \, {\pr x^\mu \over \pr \tx^\si} {\tilde J}^\si (\tx)
\, .
\eqno (2.15) $$
Corresponding to (2.12) we have
$$
{\tilde J}^0 (\tx) = e\, \de^3({\tilde \bx})\, , \qquad {\tilde {\bf J}}(\tx) = 
{\bf 0} \, .
\eqno (2.16) $$
Using (2.10) it is simple to obtain, again discarding a factor $\ep(\Omega)$,
$$
J^0 (x) = e\, (1+\ba{\cdot \bx}) \de^3 ( \bx + \half \ba (\bx^2 - t^2) ) \, , 
\qquad {\bf J}(x) = e\, \ba t \, \de^3 ( \bx + \half \ba (\bx^2 - t^2) ) \, ,
\eqno (2.17) $$
and it is straightforward to verify that this satisfies the standard
current conservation equation,
$\pr_t J^0 (x) + \bnab {\cdot {\bf J}} =0$. 

To understand further the nature of the solution (2.13) found above by 
conformal
transformation from the Coulomb electric field of a static point charge we
consider directly the solution of Maxwell's equations for an accelerating
charged particle. This is discussed in many textbooks, for a particle
moving on a trajectory $r^\mu(\tau)$, with $\tau$ the proper time, the
Li\'enard-Wiechart solution for the 4-vector potential $A^\mu=(\phi,\bA)$
takes the form
$$
A^{\mu}(x) = {e\over 4\pi}\, {{\dot r}^\mu \over |(x-r){\cdot {\dot r}}|} \, ,
\quad {\dot r}^\mu \equiv {\d r^\mu \over \d \tau} \, ,
\eqno (2.18) $$
where $\tau$ is determined by
$$
(x-r)^2 = 0 \, .
\eqno (2.19) $$
Physically it is essential to select the retarded solution so that
$t> r^0(\tau)$, as necessary for causality, but we will also consider
the advanced solution as well here. Generally it is impossible to solve
(2.19) explicitly for $\tau$ and so eliminate it from (2.18). In the case of 
interest here of a particle moving with constant acceleration $\ba$ and
passing through the origin, so that it moves on one of the branches of the
hyperbola (2.6), we may take
$$
r^\mu(\tau) = \Big ( {\sinh a\tau\over a} , {\cosh a\tau - 1\over a^2}\, \ba
\Big ) \, ,
\eqno (2.20) $$
and then (2.19) becomes
$$
(1+ \ba{\cdot \bx})(\cosh a\tau - 1) - at \sinh a\tau = \half a^2 x^2 \, .
\eqno (2.21) $$
This equation reduces to a quadratic in $e^{a\tau}$ which is easy to solve,
the solutions may be conveniently written as
$$ \eqalign {
\cosh a\tau^\pm ={}& {(1+ \ba{\cdot \bx})(1+ \ba{\cdot \bx} + \half a^2 x^2)
\mp a^2 Q t \over (1+ \ba{\cdot \bx})^2 - a^2 t^2 } \, , \cr
\sinh a\tau^\pm ={}& a \, {(1+ \ba{\cdot \bx} + \half a^2 x^2) t \mp
(1+ \ba{\cdot \bx}) Q \over (1+ \ba{\cdot \bx})^2 - a^2 t^2 } \, , \cr
Q^2 & = t^2 + (1+ \ba{\cdot \bx} + \quar a^2 x^2) x^2 =
\big (  \bx + \half \ba (\bx^2 - t^2)\big )^2 \, . \cr }
\eqno (2.22) $$
With these results 
$$
(x-r){\cdot {\dot r}} = (1+ \ba{\cdot \bx}) {\sinh a\tau^\pm \over a} - t
\cosh a\tau^\pm = \mp Q \, .
\eqno (2.23) $$
Since ${\dot r}^\mu$ is a timelike vector, ${\dot r}^0>0, \, {\dot r}^2 =-1$,
it is evident that the upper signs in (2.22,23) correspond to the usual
retarded solution while the lower to the advanced solution of Maxwell's
equations. Nevertheless it is also important to recognise that in solving
the quadratic given by (2.21) the roots should be constrained by $e^{a\tau}>0$.
Otherwise there is no point on the trajectory given by (2.20) which can
communicate with $t,\bx$. This condition restricts $t,\bx$ to the following
regions,
$$ \eqalign{
1+ \ba{\cdot \bx} + a t > {}& 0 \, , \quad \hbox{retarded solution}\, , \cr
1+ \ba{\cdot \bx} - a t > {}& 0 \, , \quad \hbox{advanced solution}\, . \cr}
\eqno (2.24) $$
Subject to this restriction on  $t,\bx$ the electromagnetic potentials
are determined from (2.18) to be
$$ \eqalign{
\bA^\pm(x) = {}& {e\over 4\pi}\, \ba \bigg ( {t\over Q}\, {1+ \ba{\cdot \bx} + 
\half a^2 x^2 \over (1+ \ba{\cdot \bx})^2 - a^2 t^2 } \mp {1+ \ba{\cdot \bx} 
\over (1+ \ba{\cdot \bx})^2 - a^2 t^2 }\bigg )  \, , \cr
\phi^\pm(x) ={}& {e\over 4\pi} \bigg ({1\over Q} (1+ \ba{\cdot \bx})
{1+ \ba{\cdot \bx} + \half a^2 x^2 \over (1+ \ba{\cdot \bx})^2 - a^2 t^2 } \mp 
{a^2 t \over (1+ \ba{\cdot \bx})^2 - a^2 t^2 } \bigg ) \, . \cr}
\eqno (2.25) $$
If we calculate the electric, magnetic fields in the usual fashion then
$$ \eqalign{
{\bf B}^\pm(x) = {}&\bnab \times \bA^\pm(x) = 
- {et \over 4\pi Q^3}\, \bx \times \ba \, , \cr
{\bf E}^\pm(x) ={}& -\bnab \phi^\pm(x) - {\pr\over \pr t} \bA^\pm(x) 
= {e\over 4\pi Q^3}
\big( \bx (1+ \ba {\cdot \bx}) - \half \ba (\bx^2 + t^2 )\big ) \, . \cr}
\eqno (2.26) $$
Remarkably the same expression results for both retarded and advanced
solutions, since the difference is expressible as a pure gauge,
$$
\bA^+ - \bA^- = - \bnab {e\over 4\pi}
\ln \big ( (1+ \ba{\cdot \bx})^2 - a^2 t^2 \big ) \, ,
\quad \phi^+ - \phi^- = {\pr \over \pr t} {e\over 4\pi}
\ln \big ( (1+ \ba{\cdot \bx})^2 - a^2 t^2 \big ) \, ,
\eqno (2.27) $$
and (2.26) is identical with the result (2.13) obtained earlier by conformal
transformation.
Of course, according to (2.24), the retarded, advanced solutions are valid in
different regions and in the space outside that given by (2.24) the solution
must be zero.

However the solution given by (2.26) with the restrictions to the regions
specified by (2.24) is not fully satisfactory since there is a discontinuity
on boundaries given by $1+ \ba{\cdot \bx} \pm a t = 0$ which leads to delta
function terms on the r.h.s. of Maxwell's equations. To resolve this paradox 
we follow the suggestion of Bondi and Gold \Bondi\ and consider, for the 
retarded solution, the more physically relevant situation where the charged
particle in the past is moving with constant velocity up to a proper time
$\tau_0$ and thereafter moves on the accelerating trajectory given by (2.20).
For a particle moving according to
$$
r_0^{\, \mu}(\tau) = c^\mu + V^\mu(\tau-\tau_0) \, ,
\eqno (2.28) $$
then the field strength is given by the relativistic formula
$$
F^{\mu\nu}(x) = {e\over 4\pi}\, {1\over R^3} \big ( V^\mu (x-c)^\nu -
V^\nu (x-c)^\mu \big ) \, , \quad R^2 = (x-c)^2 + \big ( (x-c){\cdot V}
\big )^2 \, .
\eqno (2.29) $$
For the retarded solution the trajectory (2.28) is therefore matched to (2.20) 
at $\tau = \tau_0$ by requiring
$c^\mu = r^\mu(\tau_0), \, V^\mu = {\dot r}^\mu(\tau_0)$, so that there is
no instantaneous acceleration, and we then
consider the limit $\tau_0 \to - \infty$. Writing $e^{a\tau_0} = \ep$ we
find for $\ep \to 0$ from (2.29)
$$
{\bf B}(x) \sim {e\over 4\pi R^3} \, {1\over 2\ep} \, \bx \times {\hat \ba} \, ,
\quad {\bf E}(x) \sim {e\over 4\pi R^3} \, {1\over 2\ep} \Big ( \bx_\perp
+ {{\hat \ba}\over a} ( 1+ \ba{\cdot \bx} + at ) \Big ) \, ,
\eqno (2.30) $$
where $\bx_\perp$ denotes the projection of $\bx$ perpendicular to the
acceleration $\ba$ and
$$
R^2 \sim {1\over a^2}\Big \{ {1\over 4\ep^2} \big ( 1+ \ba{\cdot \bx} + at 
- 2\ep \big )^2 + a^2 \bx_\perp {}^{\! 2} + \half \big ( 
(1+ \ba{\cdot \bx} )^2 - a^2 t^2 \big ) \Big \} \, .
\eqno (2.31) $$
This solution is valid in the region
$$
(x-c)^2 \sim {1\over a^2} \Big ( - {1\over \ep} (1+ \ba{\cdot \bx} + at )
+ 1 +  a^2 \bx_\perp {}^{\! 2} + (1+ \ba{\cdot \bx} )^2 - a^2 t^2 \Big ) > 0 
\, .
\eqno (2.32) $$
As $\ep \to 0$ $R\to \infty$ except when $1+ \ba{\cdot \bx} + at = {\rm O}(\ep)$
and we can obtain a non zero result for ${\bf B}, \, {\bf E}$
as a distribution with the aid of the result\footnote{${}^6$}{To verify this
limit we may note that if $y=(1+ \ba{\cdot \bx} + at )/2\ep$ then 
$\int_{-\infty}^{{1\over 2}
( 1 + a^2 \bx_\perp {}^{\! 2}) } \!\! \d y \, R^{-3} = {2a^3 /
( 1 + a^2 \bx_\perp {}^{\! 2} )} + {\rm O}(\ep)$.},
$$
{1\over 2\ep} \, {1\over R^3} \, \theta \big ( (x-c)^2 \big ) \sim
{2a^3 \over 1 + a^2 \bx_\perp {}^{\! 2} } \, \de ( 1+ \ba{\cdot \bx} + at) \, .
\eqno (2.33) $$
By using this in (2.30) and combining the result with (2.26) the complete
retarded solution in the limit $\ep \to 0$ becomes
$$ \eqalign {
{\bf B}^{\rm{ret}}(x) ={}& {e\over 4\pi}\bigg \{ - {t \over Q^3}\, 
\theta ( 1+ \ba{\cdot \bx} + at) + {2a^2 \over 1 + a^2 \bx_\perp {}^{\! 2} }
\, \de ( 1+ \ba{\cdot \bx} + at) \bigg \} \bx \times \ba \, , \cr
{\bf E}^{\rm{ret}}(x) ={}& {e\over 4\pi}\bigg \{  {1\over Q^3}
\big( \bx (1+ \ba {\cdot \bx}) - \half \ba (\bx^2 + t^2 )\big ) 
\theta ( 1+ \ba{\cdot \bx} + at) \cr
&  \qquad \qquad \qquad \qquad \qquad \qquad \qquad
+ {2a^2 \over 1 + a^2 \bx_\perp {}^{\! 2} }
\, \bx_\perp \, \de ( 1+ \ba{\cdot \bx} + at) \bigg \} \, . \cr}
\eqno (2.34) $$
This result is identical with Bondi and Gold \Bondi\ and it is not difficult
to check that (2.34) provides a solution of Maxwell's equations everywhere 
(when $1+ \ba{\cdot \bx} + at = 0$, $Q= (1 + a^2 \bx_\perp {}^{\! 2})/2a$). 
The analogous advanced solution is simply given in terms of (2.34) by
${\bf B}^{\rm{adv}}(\bx,t) = - {\bf B}^{\rm{ret}}(\bx,-t),\
{\bf E}^{\rm{adv}}(\bx,t) = {\bf E}^{\rm{ret}}(\bx,-t)$.

The solution exhibited in (2.13) which is non zero over all space-time 
may then be considered as a combination
of the retarded solution (2.34) together with the advanced solution for
a particle of charge $-e$ moving on the other branch of the hyperbola since
$$ \eqalign{
{\bf B}(\bx,t)={}& {\bf B}^{\rm{ret}}(\bx,t)- 
{\bf B}^{\rm{adv}}(-\bx-2\ba/a^2,t) \, , \cr
{\bf E}(\bx,t)={}& {\bf E}^{\rm{ret}}(\bx,t)+ 
{\bf E}^{\rm{adv}}(-\bx-2\ba/a^2,t) \, , \cr}
\eqno (2.35) $$
which has no discontinuities.
\bigskip
\noindent{\bigbf 3 Projective Cone}
\medskip
The action of conformal transformations on Minkowski space is clearly non
linear but a significant simplification is possible by introducing an
associated space on which the conformal group acts linearly. To understand
this we consider a general infinitesimal transformation for which
$$
x^\mu = \tx^\mu + v^\mu (\tx) \, , \qquad \Omega(\tx) = 1 - \si (\tx) \, ,
\eqno (3.1) $$
so that (1) requires
$$
\pr_\mu v_\nu + \pr_\nu v_\mu =  2 \si g_{\mu\nu} \, .
\eqno (3.2) $$
This has the general solution
$$
v^\mu (x) = a^\mu + \omega^\mu{}_\nu x^\nu + c x^\mu +
b^\mu x^2 - 2 x^\mu b {\cdot x} \, , \quad  \omega_{\mu\nu} = -
\omega_{\nu\mu} \, , \quad \si(x) = c  - 2 b {\cdot x} \, .
\eqno (3.3) $$
The 15 parameters determining $v$ may be written as a $6\times 6$ matrix,
$$
W^A{}_B = \pmatrix{\omega^\mu{}_\nu & a^\mu - b^\mu & a^\mu+b^\mu\cr
-a_\nu + b_\nu & 0 & -c \cr a_\nu + b_\nu & - c & 0\cr}\, ,
\eqno (3.4) $$
where $A=0,1,2,3,5,6$. The matrix $W$ is so defined that it provides a 
representation of the Lie algebra of conformal transformations
$$
v'{}^\mu = v_2{\cdot \pr}\, v_1{}^{\! \mu} -  v_1{\cdot \pr}\, v_2{}^{\! \mu}
\ \Rightarrow \ W' = [W_1 ,W_2] \, .
\eqno (3.5) $$
Furthermore
$$ W_{AB} = g_{AC} W^C{}_B = - W_{BA} \, , \quad g_{AB} = {\rm diag.}
(-1,1,1,1,1,-1) \, ,
\eqno (3.6) $$
so that it represents a generator of the group $SO(4,2)$.

To describe the standard conformal compactification of Minkowski space we 
introduce the equivalence class
of real six dimensional vectors $\eta^A=(\eta^\mu,\eta^5,\eta^6)$  denoted
by $[\eta^A]$ and defined by
$$
g_{AB} \eta^A \eta^B = 0 \, , \qquad \eta^A \sim \lambda \eta^A \, , \quad
\lambda \ne 0 \, .
\eqno (3.7) $$
$[\eta^A]$ specifies a point on a four dimensional projective cone.
Clearly the conditions (3.7) are invariant under $\de \eta^A = W^A{}_B \eta^B$
so that there is a natural action of $SO(4,2)$ on this space.
By suitable choice of $\lambda>0$ the constraint $g_{AB} \eta^A \eta^B = 0$
can be re-written as
$$
{\bold \eta}^2 + (\eta^5)^2 = (\eta^0)^2 + (\eta^6)^2 = 1 \, ,
\eqno (3.8) $$
so that the projective cone may be identified with $S^3 \times S^1/Z_2$ with 
$Z_2$ corresponding to the identification under the reflection 
$\eta^A \to - \eta^A$, or taking $\lambda = -1$ in (3.7). 
It is convenient to define
$$
\eta^{\pm} = \eta^6 \pm \eta^5
\eqno (3.9) $$
and then for any $\eta^+ \ne 0$ we may define a point $x^\mu \in M^4$, four
dimensional Minkowski space, by
$$
x^\mu = {\eta^\mu \over \eta^+} \, .
\eqno (3.10) $$
Using $\de\eta^\mu = \omega^\mu{}_\nu \eta^\nu + a^\mu \eta^+ + b^\mu \eta^-, \,
\de \eta^+ = 2b_\nu \eta^\nu - c \eta^+ , \, \eta^- = x^2 \eta^+$ it is easy
to see that this is compatible with $\de x^\mu = v^\mu(x)$ as given by (3.3).
An inversion through $x=0$ corresponds to $\eta^5 \to - \eta^5$ or
$\eta^+ \leftrightarrow \eta^-$.
Conversely for any $x^\mu \in M^4$ we may define a unique vector, within
the equivalence class defined by (3.7), satisfying $\eta^+=1$, by
$$
\eta^A(x) = \big ( x^\mu, \half(1-x^2), \half(1+x^2) \big) \, .
\eqno (3.11) $$
For $\eta^+=0$ a representative of the equivalence class given by (3.7) may be 
expressed by
$$
\eta_\infty^{\, A}(n) = \big ( n^\mu, -\half, \half \big ) \, , 
\quad n^2=0 \, ,
\eqno (3.12) $$
for any null vector $n$. The set of such null vectors defines a 
three dimensional light cone
$C^3$. It is not difficult to see that the limits of $\eta^A(x)$ as the point 
$x$ tends to $\infty$ along the direction defined by a null 4-vector $n^\mu$
are realised by $\eta_\infty^{\, A}(\lambda n)$ for some non zero $\lambda$.
If $t\to \pm \infty$ or $|\bx|\to \infty$ then the limit is given by the
single point $\eta_\infty^{\, A}(0)$. Thus by including its limiting points
in this fashion Minkowski space is compactified to
$\overline {M^4} = M^4 \cup \{ C^3\}_\infty  \simeq S^3\times S^2/Z_2$.

{}From (3.3) and (3.4) we may easily verify that
$$
\big ( v(x) {\cdot \pr} - \si (x) \big ) \eta^A(x) = W^A{}_B \eta^B(x) \, ,
\eqno (3.13) $$
which demonstrates again the relation between the non linear action of the
conformal group on $M^4$ to its linear realisation on the projective cone.
It is also useful to note that
$$
\eta^A(x) = g^A{}_B (x) \eta^B(0) \ \ \hbox{or} \ \
\eta_A(x) g^A{}_B (x) = \eta_B(0) \, ,
\eqno (3.14) $$
where $g^A{}_B (x)$ belongs to $SO(4,2)$ and is given explicitly by
$$
g^A{}_B (x) = \pmatrix{\de^\mu{}_{\!\nu} & x^\mu & x^\mu \cr
- x_\nu & 1-\half x^2 & -\half x^2 \cr x_\nu & \half x^2 & 1+\half x^2 \cr } 
\, .
\eqno (3.15) $$
For any homogeneous function on the projective cone, $f(\lambda \eta) =
\lambda^{-y} f(\eta)$, it follows from (3.13) that
$$
\big ( v(x) {\cdot \pr} +y\, \si (x) \big ) f(\eta(x)) = - W^{AB} L_{AB}
f(\eta) \big |_{\eta =\eta(x)} \, ,
\eqno (3.16) $$
with
$$
L_{AB} = \eta_A {\pr \over \pr \eta^B} - \eta_B {\pr \over \pr \eta^A} \, ,
\eqno (3.17) $$
the generators of the $SO(4,2)$ Lie algebra
$$
[L_{AB},\, L_{CD}] = g_{BC} L_{AD} - g_{AC} L_{BD} - g_{BD} L_{AC} +
g_{AD} L_{BC} \, .
\eqno (3.18)$$ 
The result (3.17) shows how conformal transformations on scalar fields on $M^4$
become essentially linear rotations acting on the corresponding fields
defined on the projective cone. The analogous result for tensor fields
depends on
$$
v(x) {\cdot \pr}\, g^A{}_\mu (x) - g^{A\nu}(x) \pr_{[\nu}v_{\mu]}(x)
= W^A{}_B g^B{}_\mu ( x) - 2b_\mu \eta^A(x) \, ,
\eqno (3.19) $$
which may be obtained from (3.13) using $\pr_\mu \eta^A(x) = g^A{}_\mu (x)$.
Thus $g^A{}_\mu (x)$ provides the necessary link between covariant tensors
on the two spaces although the last term in (3.19), involving $b_\mu$, spoils
the required relation between the transformation rules and so this term
must separately vanish. 

If we apply the above discussion of the compactification of Minkowski space
to the trajectory of an accelerating particle defined by (2.16) then using
(3.11)
$$
\eta^A\big (r(\tau)\big ) \sim 
- {1\over a^2} e^{\pm a\tau}\big ( n_{\pm}{}^{\!\mu},
-\half, \half \big ) \, , \quad n_{\pm}{}^{\!\mu} 
= -\half a(\pm 1, {\hat \ba}) \, , \ \hbox {as} \ \tau \to \pm \infty \, .
\eqno (3.20) $$
Correspondingly on the other branch of the hyperbola which may be parameterised
by
$$
{\bar r}^\mu(\tau) = \Big ( {\sinh a\tau\over a} , - {\cosh a\tau + 1\over a^2}
\, \ba \Big ) \, ,
\eqno (3.21) $$
the limits are
$$
\eta^A\big ({\bar r}(\tau)\big ) \sim {1\over a^2} e^{\pm a\tau}
\big ( n_{\mp}{}^{\!\mu}, -\half, \half \big ) \, , \ \hbox {as} \ 
\tau \to \pm \infty \, .
\eqno (3.22) $$
Clearly the two branches of the hyperbola form a closed curve on
$\overline {M^4}$, since from (3.20,22)
$[\eta^A(r(\pm \infty))] = [\eta^A({\bar r}(\mp \infty))]$, which correspond
to the line ${\bold \eta}= - \half \ba \, \eta^-$ on the projective cone.

To illustrate this we may restrict $\ba$ to the 1-direction and then, imposing
(3.8), write $\eta^1 = \sin \theta$ and $\eta^0 = \cos {\hat \tau}$, with
${\hat \tau}$ a conformal time, where we identify $(\theta,{\hat \tau})$ and 
$(\theta\pm\pi,{\hat \tau}\pm\pi)$. The equation for hyperbolic motion in these
variables becomes $\cos {\hat \tau} = \cos \theta - {2\over a} \sin \theta$,
which is plotted on the appropriate Penrose diagram in figure 1.
\medskip

{\baselineskip 10pt
\line{\vtop{\hsize=4.45cm
{\noindent{\eightpoint Figure 1. 
Penrose diagram for conformally compactified Minkowski space
$\overline{M^4}$ describing the trajectory of a particle
moving with constant acceleration.
$\overline{M^4}$ is represented in the figure by the
diamond region with opposite sides identified, so that $\iota^\pm,\iota^0$
denote a single point. On $\overline{M^4}$ $A,A'$ and $B,B'$ are identified
so that the two branches of the hyperbolic motion form a single closed curve.}}}
\hfil\vtop{\null\vskip-0.5cm\epsfxsize=0.6\hsize\epsfbox{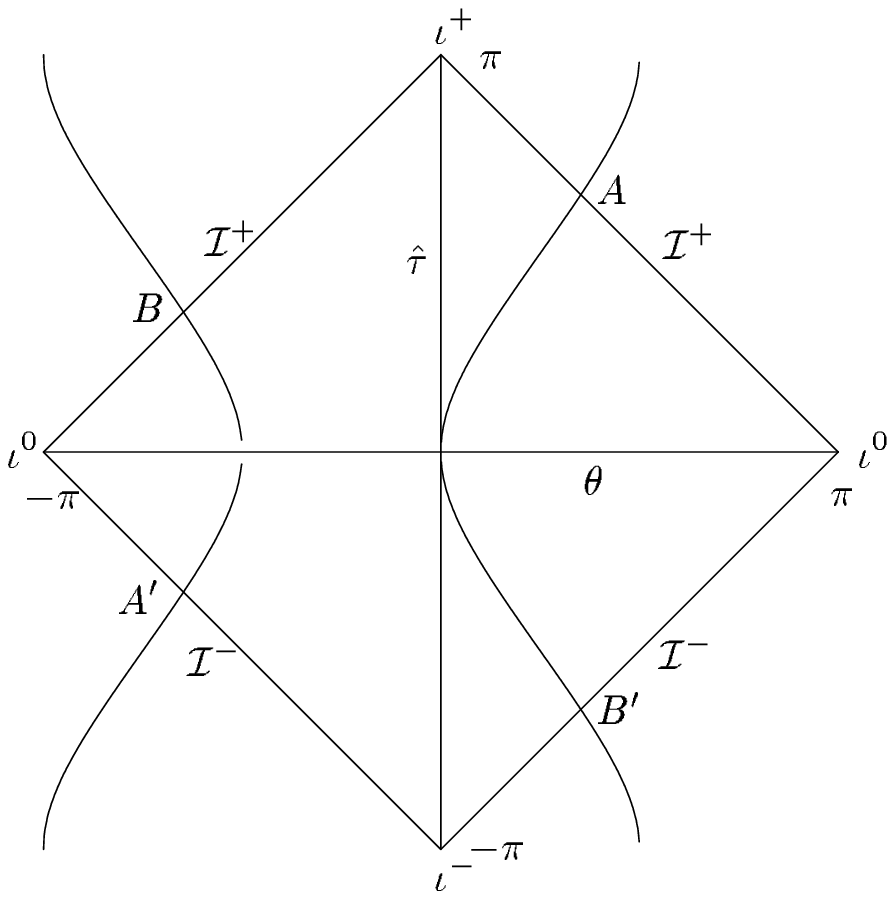}}}}
\medskip
\noindent{\bigbf 4 Electrodynamics on the Projective Cone}
\medskip
Dirac \Dirac\ was apparently the first to realise that it would be natural to
reformulate conformal field theories on the projective
cone given by (3.7) in terms of fields which are homogeneous functions of
$\eta$ so that conformal invariance would become explicit. For electrodynamics
it is appropriate to introduce a gauge field $\A_A(\eta)$ satisfying
$$
\A_A(\lambda\eta) = \lambda^{-1}\A_A(\eta) \, , \qquad \A_A(\eta) \eta^A =0\, ,
\eqno (4.1) $$  
and which has an additional gauge freedom
$$
\A_A(\eta) \sim \A_A(\eta) + \eta_A \P(\eta) \, .
\eqno (4.2) $$
The relation to the standard electromagnetic 4-vector gauge potential on 
Minkowski space is given by
$$
A_{\mu}(x) = \A_A(\eta(x)) g^A{}_{\mu}(x) = \A_\mu (\eta(x)) + 2x_\mu
\A_-(\eta(x)) \, , \quad \A_\pm = \half (\A_6 \pm \A_5)  \, ,
\eqno (4.3) $$
with $\eta(x)$ given by (3.11) and $g^A{}_{\mu}(x)$ specified by (3.15).
Since $\eta_A(x)g^A{}_\mu(x) = 0$ it is clear that $A_\mu$ is invariant under
(4.2). Alternatively (4.3) may be written more succinctly as
$$
A_\mu(x) \d x^\mu = \A_A(\eta(x)) \d \eta^A(x) \, .
\eqno (4.4) $$

In a similar fashion the field strength $F_{\mu\nu}$ is related to $\F_{AB}$ by
$$
F_{\mu\nu}(x) = \F_{AB}(\eta(x)) g^A{}_{\mu}(x) g^B{}_{\nu}(x) \, ,
\eqno (4.5) $$
where
$$
F_{\mu\nu} = \pr_\mu A_\nu - \pr_\nu A_\mu \, , \qquad
\F_{AB}(\eta) = {\pr\over \pr \eta^A} \A_B(\eta) -
{\pr\over \pr \eta^B} \A_A(\eta) \, , 
\eqno (4.6) $$
and $\F_{AB}$ has the homogeneity properties
$$
\quad \F_{AB}(\lambda\eta) = \lambda^{-2} \F_{AB}(\eta) \, .
\eqno (4.7) $$
The condition $\A_A(\eta)\eta^A = 0$ may be seen from the result (3.19) to
be necessary if $\A_A$ is a vector field under $SO(4,2)$ and  corresponds
to an $A_\mu$ which transforms irreducibly as a so-called quasi-primary field.
For the corresponding condition on $\F_{AB}$ to be an anti-symmetric tensor
field corresponding to a quasi-primary $F_{\mu\nu}$ it is sufficient to require,
on the projective cone,
$$
\F_{AB} (\eta)\eta^B = \eta_A \Y(\eta) \, .
\eqno (4.8) $$

However it is important to recognise that the derivative $\pr_A$ cannot in
general be regarded as a well defined operation acting on fields over
the projective cone determined by the
relations (3.7) since, if all components $\eta^A$ are taken as independent,
$\pr_A$ does not commute with the constraint $\eta^2 \equiv g_{AB}\eta^A\eta^B 
= 0$. In consequence $\pr_A$ acting on $f(\eta)$, which is specified for
$\eta^2=0$, is undetermined up to terms proportional
to $\eta_A$ since taking $f(\eta)\sim f(\eta) + \half \eta^2 g(\eta)$, with
$g(\eta)$ arbitrary, then $\pr_A  f(\eta) \sim \pr_A  f(\eta) + \eta_A g(\eta)$
after setting $\eta^2=0$. However the generators $L_{AB}$ given by (3.17)  
and $\eta^A\pr_A$ satisfy $[L_{AB},\eta^2]=0$ and 
$\eta^A\pr_A\, \eta^2 = \eta^2(\eta^A\pr_A+2)$ respectively and are 
therefore well defined intrinsic operators acting on fields over 
the projective cone. Obviously for homogeneous fields $\eta^A\pr_A$ can be 
replaced by the homogeneity degree. As a result of the arbitrariness in the 
action of $\pr_A$ there remains a gauge freedom in the field
strength $\F_{AB}$, initially introduced in (4.6), which is expressible as
$$
\F_{AB} (\eta) \sim \F_{AB} (\eta) + \eta_A \B_B(\eta) - \eta_B \B_A (\eta)\, ,
\eqno (4.9) $$
with $\B_A(\eta)$ homogeneous of degree $-3$ and otherwise arbitrary.
Clearly such arbitrariness leaves $F_{\mu\nu}(x)$ in (4.5) invariant while
from (4.8)
$$
\Y(\eta) \sim \Y(\eta) + \B_A(\eta) \eta^A \, .
\eqno (4.10) $$
It is crucial that any re-writing Maxwell's equations in terms of $\F_{AB}$
should be invariant in form under variations given by (4.9).

We discuss first the Bianchi identity which is now assumed to be expressed, 
by allowing for the arbitrariness in the definition of the derivatives, as
$$ \eqalign{
3\pr_{[A} \F_{BC]} \equiv {} & \pr_A \F_{BC} + \pr_B \F_{CA} + \pr_C \F_{AB} \cr
={} & \eta_A \X_{BC} + \eta_B \X_{CA} + \eta_C \X_{AB} \, , \qquad 
\X_{AB}=-\X_{BA}\, ,  \cr}
\eqno (4.11) $$
with $\X_{AB}(\eta)$ homogeneous of degree $-4$.
Extending the results in (4.5,6) it is easy to see that the freedom
allowed on the r.h.s. of (4.11) does not affect the usual Maxwell equation
$\pr_{[\lambda} F_{\mu\nu]} = 0$. A corollary of (4.11) is
$$ \eqalign{
12 & \eta_{[A} \pr_B \F_{CD]} \cr
& = L_{AB} \F_{CD} - L_{AC} \F_{BD} + L_{AD} \F_{BC}
+ L_{BC} \F_{AD} - L_{BD}\F_{AC} + L_{CD} \F_{AB} = 0 \, , \cr}
\eqno (4.12) $$
which only involves the unambiguous generator $L_{AB}$ defined in (3.17).

To see the necessity in general of allowing for a non zero r.h.s. in 
(4.11) we first note that the arbitrariness in the action of derivatives
does not change the standard result,
$$
[\pr_A, \, \eta_B] = g_{AB} \, ,
\eqno (4.13)
$$
since this commutes with $\eta^2$. Applying this with (4.9) then gives
$$
\pr_C \F_{AB} \sim \pr_C \F_{AB} + g_{CA}\B_B - g_{CB}\B_A + \eta_A \pr_C
\B_B - \eta_B \pr_C \B_A  \, .
\eqno (4.14) $$
Using this in the l.h.s of (4.11) it is straightforward to see that (4.14) is
compatible with the r.h.s. of (4.11) if we also take
$$
\X_{AB} \sim \X_{AB} - (\pr_A \B_B - \pr_B \B_A )
+ \eta_A \W_B  - \eta_B \W_A   \, .
\eqno (4.15) $$
The freedom of terms involving $\W_A$ is introduced since they
cancel identically on the r.h.s. of (4.11) and so
represent an additional gauge freedom in the precise form of $\X_{AB}$. For
later use we note that
$$
\X_{AB}\eta^B \sim \X_{AB}\eta^B - 2\B_A - \pr_A (\B{\cdot \eta})
+ \eta_A \, \W{\cdot \eta} \, .
\eqno (4.16) $$

In order to discuss the remaining Maxwell equation, which relates the fields
to the electromagnetic current, it is convenient first to rewrite (4.5) as 
$$
F^{\mu\nu}(x) = g^{-1\, \mu}{}_{\! A}(x) g^{-1\, \nu}{}_{\! B}(x)
\F^{AB}(\eta(x))\, .
\eqno (4.17) $$
Regarding $\F^{AB}(\eta)$ as defined for arbitrary $\eta^A$ we may then obtain
$$ \eqalign {
\pr_\nu F^{\mu\nu}(x) = g^{-1\, \mu}{}_{\! A}(x)\bigg \{&{\pr \over \pr \eta^B}
\F^{AB}(\eta) \cr
+ {2\over \eta^+}\, & {\pr \over \pr \eta^-} \big (\eta_B 
\F^{AB}(\eta) \big ) - {1\over \eta^+}\Big ( 2 + \eta^B {\pr \over \pr \eta^B}
\Big ) \F^{A+}(\eta)  \bigg \} \bigg |_{\eta =\eta(x)} \, . \cr}
\eqno (4.18) $$
The last term in (4.18) vanishes if $\F^{AB}$ satisfies the homogeneity 
condition (4.7) and then it can be re-written, using $\pr_- \eta^2 = - \eta^+$,
as
$$ 
\pr_\nu F^{\mu\nu}(x) = g^{-1\, \mu}{}_{\! A}(x)\bigg \{{\pr \over \pr \eta^B}
\F^{AB}(\eta) - \eta_B \X^{AB}(\eta)- \pr^A \Y(\eta) + {2\over \eta^+}\, 
{\pr \over \pr \eta^-} \Q^A(\eta) \bigg \} \bigg |_{\eta =\eta(x)} \, ,
\eqno (4.19) $$
where
$$
\Q^A(\eta) = \eta_B \F^{AB}(\eta) - \half \eta^2 \big ( \eta_B \X^{AB}(\eta)
+ \pr^A \Y(\eta) \big ) \, .
\eqno (4.20) $$
In an appendix we show that $\Y(\eta)$ can be chosen so that 
$\Q^A(\eta) = \eta^A \Y(\eta) + \rO((\eta^2)^2)$, which reduces to (4.8)
if $\eta^2=0$, and hence this term gives a zero contribution on the r.h.s. 
of (4.19) as a consequence of $g^{-1\, \mu}{}_{\! A}(x) 
\eta^A(x) = g^{-1\, \mu}{}_{\! -}(x) =0$. The usual Maxwell equation
relating the fields to the charge current density, $\pr_\nu F^{\mu\nu}=J^\mu$,
can then be written in a manifestly conformally invariant form
$$
\pr_B \F^{AB}(\eta) - \eta_B \X^{AB}(\eta) - \pr^A \Y(\eta) = \J^A(\eta) \, ,
\eqno (4.21) $$
where we take
$$
J^\mu(x) = g^{-1\, \mu}{}_{\! A}(x) \J^A(\eta(x)) \, ,
\eqno (4.22) $$
for $\J^A(\eta)$, homogeneous of degree $-3$, the current density on the 
projective cone. From (3.19) for $\J^A$ to transform as a vector
it is necessary to impose the condition
$$
\eta_A \J^A(\eta) = 0 \ \ \hbox{for} \ \ \eta^2 =0 \, ,
\eqno (4.23) $$
and also in (4.22) $\J^A$ is arbitrary up to
$$
\J^A(\eta) \sim \J^A(\eta) + \eta^A \K(\eta) \, .
\eqno (4.24) $$

An important check on the equation (4.21) is that it respects the requirement
of invariance under the variations given by (4.9) since from (4.14) we have
$$
\pr_B \F^{AB} \sim \pr_B \F^{AB} - 2\B^A + \eta^A \, \pr {\cdot \B}  \, ,
\eqno (4.25) $$
and using (4.10) and (4.16) the variation may be absorbed in the freedom of
the current exhibited in (4.24).

Just as (4.19) was obtained we may derive similarly from (4.22) an equation
related to the conservation of the electromagnetic current,
$$ 
\pr_\mu J^{\mu}(x) = \bigg \{{\pr \over \pr \eta^A} \J^A(\eta)
+ {2\over \eta^+}\, {\pr \over \pr \eta^-} \big (\eta_A \J^A(\eta) \big )
- {1\over \eta^+}\Big ( 3 + \eta^B {\pr \over \pr \eta^B}
\Big ) \J^{+}(\eta)  \bigg \} \bigg |_{\eta =\eta(x)} \, . 
\eqno (4.26) $$
The last term in (4.26) vanishes since $\J^{A}$ is homogeneous of degree $-3$
and then the conservation equation $\pr_\mu J^{\mu}=0$ is satisfied if
$$
\eta_A \J^A(\eta) = \half \eta^2 \L(\eta) \, , \qquad 
\big ( \pr_A \J^A - \L \big ) \big |_{\eta^2=0} = 0 \, .
\eqno (4.27) $$
Note that under the variation (4.24) $\L \sim \L + 2 \K$ while 
$ \pr_A \J^A \sim \pr_A \J^A + 2 \K$, since $\K(\eta)$ has degree $-4$, so
that (4.27) is invariant while it is also obviously compatible with (4.23).

It is now feasible to re-express the above field equations so that they
only involve intrinsic differential operators, such as $L_{AB}$ in (3.17),
which are well defined on the projective cone. The current conservation
equations in (4.27) can be used to obtain \consv\
$$ \eqalignno{
L_{AB} \J^B ={}& \eta_A \pr_B \J^B + \J_A - \pr_A (\eta_B \J^B ) \cr
={}& \J_A \ \ \hbox{for} \ \ \eta^2 =0 \, . & (4.28) \cr}
$$
The version of Maxwell's equation relating the fields to the current given
by (4.21) may also be re-written with the aid of the result
$$
\eta^A \pr_C \F^{BC} - \eta^B \pr_C \F^{AC} =  L^A{}_C  \F^{BC} 
- L^B{}_C \F^{AC} +\eta_C \, 3\pr^{[A} \F^{BC]} - \eta{\cdot \pr} \F^{AB} \, ,
\eqno (4.29) $$
in the more natural form
$$
L^{[A}{}_C \F^{B]C} + \F^{AB} - \half L^{AB} \Y = \eta^{[A} \J^{B]} \, .
\eqno (4.30) $$
which again only involves the operators $L_{AB}$. Obviously the r.h.s. of 
(4.30) is invariant under variations as in (4.24) although the invariance 
under (4.9) is now less evident.

{}From the essential equation (4.30) necessary conditions on the current
$\J^A$ follow just as the usual current conservation equation must be imposed
for consistency of Maxwell's equations. Using
$$ \eqalign {
\eta_B \big ( & L^A{}_C \F^{BC} - L^B{}_C \F^{AC} + 2 \F^{AB} - L^{AB} \Y \big )
\cr
= {}& L^A{}_C (\eta_B \F^{BC} ) + 3 \eta_B \F^{AB} + \eta_C\, \eta{\cdot \pr}
\F^{AC} -\eta^A \, \eta{\cdot \pr} \Y \cr
= {}& - L^A{}_C ( \eta^C \Y) + 3\eta^A \Y 
=  - \eta^A ( 2 + \eta{\cdot \pr})\Y = 0 \, , \cr}
\eqno (4.31) $$
it is clear that we must require (4.23).
By contracting (4.21) with $\eta_A$ this result leads to
$$
L_{AB} \F^{AB} = - 4\Y \, .
\eqno (4.32) $$
Other consistency conditions which flow from (4.30) are discussed in the
appendix.

As we have made clear due to the arbitrariness of $\F_{AB}$ under variations of
the form (4.9) it does not enjoy the same invariant status as the usual
field strength tensor $F_{\mu\nu}$ on Minkowski space. However we may define
$$ \eqalignno{
{\tilde \F}_{ABC} = {}&\eta_A \F_{BC} +  \eta_B \F_{CA} +  \eta_C \F_{AB} 
&(4.33a) \cr
= {}& L_{AB} \A_C + L_{BC} \A_A + L_{CA} \A_B \, , & (4.33b) \cr}
$$
which is obviously invariant. Clearly from (4.7) ${\tilde \F}_{ABC}(\eta)$ is
homogeneous of degree $-1$. It is also evident from (4.33a) and (4.8)
that this antisymmetric rank 3 tensor satisfies\footnote{${}^7$}{From (4.33b)
we may also write ${\tilde \F}_{ABC}\eta^C = L_{AB}(\A_C\eta^C) - \eta_A
(\eta{\cdot \pr}+1) \A_B + \eta_B (\eta{\cdot \pr}+1) \A_A$, which shows
how both conditions in (4.1) are necessary for this to be zero.} 
$$
{\tilde \F}_{[ABC}(\eta) \eta_{D]} = 0 \, , \qquad {\tilde \F}_{ABC}(\eta) 
\eta^C = 0 \ \ \hbox{so long as} \ \ \eta^2 =0 \, .
\eqno (4.34) $$
Subject to these conditions it is easy to see that ${\tilde \F}_{ABC}$ has
6 degrees of freedom, just like $F_{\mu\nu}$ which can be directly
expressed in terms of ${\tilde \F}_{ABC}$ by
$$
{\tilde \F}_{ABC}(\eta(x))g^A{}_-(x)g^B{}_\mu(x)g^C{}_\nu(x) = - \half
F_{\mu\nu}(x)\, .
\eqno (4.35) $$
It is also possible to re-write
the basic equation (4.30) in terms of ${\tilde \F}_{ABC}$ since (4.30) is
equivalent to
$$
\nab_C {\tilde \F}^{ABC} = \eta^A \J^B - \eta^B \J^A \, ,
\eqno (4.36) $$
with $\nab_C$ a differential operator, essentially introduced by Binegar, 
Fronsdal and Heidenreich \Fron, which is defined in this case by
$$
\nab_C {\tilde \F}^{ABC} \equiv \pr_C {\tilde \F}^{ABC} - \half \pr^2\big(
\eta_C {\tilde \F}^{ABC} \big ) \, .
\eqno (4.37) $$
The action of $\nab$ in (4.37) is assumed to be calculated for $\eta^A$
unconstrained by $\eta^2=0$.
It is then clear that this operator has the essential property 
$\nab_C \big (\eta^2 \G^{ABC} \big) \propto \eta^2$, for any $\G^{ABC}(\eta)$
which is homogeneous of degree $-3$. Hence there are no ambiguities in
imposing  $\eta^2=0$ and  $\nab$ is an intrinsic operator on
the projective cone.\footnote{${}^8$}{If space-time has an arbitrary dimension 
$d$ and for a vector field $\O^A(\eta)$ on the associated projective cone,
homogeneous of degree 
$-y$, the general definition of the divergence of $\O^A$ becomes
$\nab_A\O^A \equiv \pr_A \O^A + \pr^2 \big ( \eta_A \O^A )/(2y-d)$ so long as
$y\ne {1\over 2} d$. This makes it clear that $\nab_B \F^{AB}$ is undefined
when $d=4$ and $y=2$.} To show the equivalence of (4.37) with (4.30) it is
first convenient to write
$$
\pr_C {\tilde \F}^{ABC} = 4 \F^{AB} + L^A{}_C \F^{BC} -  L^B{}_C \F^{AC}
+ \pr^A(\eta_C \F^{BC} ) - \pr^B(\eta_C \F^{AC} ) \, .
\eqno (4.38) $$
We then extend (4.8) to $\eta^2 \ne 0$ by writing\footnote{${}^9$}{Using this
result we may also obtain (4.32) more directly since $L_{AB}\F^{AB}
=- 2\pr_A(\eta_B\F^{AB}) \allowbreak = - 2\pr_A (\eta^A \Y + \eta^2 \U^A) 
= - 4\Y$ after setting $\eta^2=0$.}
$$
\eta_C \F^{AC} = \eta^A \Y + \eta^2 \U^A \, , \quad \eta_A \U^A = - \Y \, ,
\eqno (4.39) $$
so that
$$
\big (\pr^A(\eta_C \F^{BC} )- \pr^B(\eta_C \F^{AC} ) \big ) \big |_{\eta^2 = 0}
= - L^{AB} \Y + 2(\eta^A \U^B - \eta^B \U^A ) \, . 
\eqno (4.40) $$
With the aid of (4.39) and (4.33a) it is easy to see that
$$
\eta_C {\tilde \F}^{ABC} = \eta^2 ( \F^{AB} + \eta^A \U^B - \eta^B \U^A ) \, .
\eqno (4.41) $$
In general
$$
\pr^2 (\eta^2 X) \big |_{\eta^2 = 0} = 12 X + 4 \eta{\cdot \pr} X \, ,
\eqno (4.42) $$
so that $\pr^2\big( \eta_C {\tilde \F}^{ABC} \big )$ can be immediately
determined giving, in combination with (4.36,40),
$$
\nab_C {\tilde \F}^{ABC} = 2 \F^{AB} + L^A{}_C \F^{BC} -  L^B{}_C \F^{AC}
- L^{AB} \Y \, ,
\eqno (4.43) $$
as required. It is also of interest to note that the conservation equation
(4.27) may be written as
$$
\nab_A \J^A \equiv \pr_A  \J^A + \half \pr^2 ( \eta_A \J^A) = 0 \ \
\hbox{for} \ \ \eta^2=0 \, .
\eqno (4.44) $$

The equations (4.30) or (4.36) describing electrodynamics are manifestly
invariant under the conformal group $O(4,2)$ but it is important to stress
that their justification is only valid in four space-time dimensions, in
accord with the standard conclusions for conformal invariance of Maxwell's
equations on Minkowski space.
\bigskip
\noindent{\bigbf 5 Elementary Solutions on the Projective Cone}
\medskip
The transformation of Maxwell's equations for classical electrodynamics
to the projective cone allows us to find elementary solutions which are
equivalent to (2.13) under the relations (4.4) or (4.17). The solution
may be written as
$$
\F_{AB}(\eta) = {e\over 4\pi}\, {1\over | {\bold \eta}+ \half \ba \,\eta^- |^3}
\, \T_{ABC}\eta^C \, ,
\eqno (5.1) $$
with $\T_{ABC}=\T_{[ABC]}$ totally antisymmetric and given by
$$
\T_{0+-} = \half \, , \qquad \T_{i0-} = -\half a_i \, .
\eqno (5.2) $$
In this case the action of conformal transformations is very simple, if
${\tilde \F}_{AB}(\eta)$ denotes the solution for $\ba={\bold 0}$, then
$$
\F_{AB}(\eta) = {\tilde\F}_{CD}(G\eta) G^C{}_A G^D{}_B \ \ \hbox{or} \ \
\T_{ABC} = {\tilde \T}_{DEF}G^D{}_A G^E{}_B G^F{}_C \, ,
\eqno (5.3) $$
where $G^A{}_B$ is a matrix belonging to $SO(4,2)$ defined by the non
zero elements,
$$
G^0{}_0 = G^+{}_{\! +} = G^-{}_{\! -} = 1 \, , \quad G^i{}_j = \de^i{}_{j} \, ,
\quad G^i{}_- = \half a_i \, , \quad G^+{}_j = a_j \, , \quad
G^+{}_{\! -} = \quar \ba^2 \, .
\eqno (5.4) $$

{}From the explicit expression (5.1) it is trivial to see that, with the
definition (4.8), in this case
$$
\Y(\eta) = 0 \, ,
\eqno (5.5) $$
and with more effort from (4.11),
$$
\X_{AB}(\eta) = - {e\over 4\pi}\, 
{3\over | {\bold \eta}+ \half \ba \,\eta^- |^5} \, \T_{ABC}\eta^C \, .
\eqno (5.6) $$
From (5.1,6) it is easy to see that
$$
\pr_B \F^{AB} = 0 \, , \qquad \X_{AB} \eta^B = 0 \, ,
\eqno (5.7) $$
so that, with (5.5), the equation (4.21) is satisfied, at least for
$  {\bold \eta} \ne - \half \ba \,\eta^-$.

In order to obtain an expression for the current density on the projective
cone it is necessary to pay more attention to the singularity as
$ {\bold \eta} \to - \half \ba \,\eta^-$. The result (5.1) is non integrable
on a one dimensional subspace of the projective cone so we introduce a
regularisation by replacing $|{\bold \eta} + \half \ba \,\eta^-|$ by $\R$
where
$$
\R^2 = ({\bold \eta} + \half \ba \,\eta^-)^2 + \ep (\eta^{+})^2 \, , \quad
\ep >0 \, ,
\eqno (5.8) $$
and then take the limit $\ep\to 0$. With this definition $\R^{-3}$ is
integrable and we may straightforwardly verify that
$$
\eta^+ {\pr\over \pr \eta^+} {1\over \R^3} = - 3 \, {\ep (\eta^{+})^2\over
\R^5} \sim - 4\pi\, \de^3({\bold \eta} + \half \ba \,\eta^-) \ \ \hbox{as} \ \
\ep \to 0 \, .
\eqno (5.9) $$
It is trivial to see that
$ L^A{}_C \T^{BC}{}_{\! D}\eta^D + \T^{AB}{}_{\! C}\eta^C = 0 $
and hence 
$$
L^{[A}{}_C \F^{B]C}(\eta) + \F^{AB}(\eta) \sim {e\over 4\pi} L^{[A}{}_C 
{1\over \R^3} \, \T^{B]C}{}_{\! D} \eta^D   \ \ \hbox{for} \ \ \ep \to 0 \, .
\eqno (5.10) $$
Using (5.5) the field equation (4.30), with the aid of (5.9), then gives
a non zero r.h.s. as a consequence of using (5.9) for the action of
$\pr/\pr \eta^+$ so that we can now obtain
$$
\eta^{[A} \J^{B]}(\eta) = e \, \half \T^{AB}{}_{\! C}\eta^C \,
\de^3({\bold \eta} + \half \ba \,\eta^-) \, .
\eqno (5.11) $$
Since
$$
J^\mu(x) = 2 g^{-1\,+}{}_{\! A}(x) g^{-1\, \mu}{}_{\! B}(x) \,\eta^{[A}(x)
\J^{B]}(\eta(x)) \, ,
\eqno (5.12) $$
it is easy to see that this is in agreement with (2.16).

Alternatively we may use this regularisation to evaluate $\pr_A \F_{BC}$
and in (4.11) we then find, instead of (5.6),
$$
\X_{AB}(\eta) = - {e\over 4\pi}\, 
{3\over \R^5} \, \T_{ABC}\eta^C  + {\hat \X}_{AB} (\eta) \, ,
\eqno (5.13) $$
where ${\hat \X}_{AB} (\eta)\propto \de^3({\bold \eta} + \half \ba \,\eta^-)$
and, using (5.9), has the explicit form
$$ \eqalign {
{\hat \X}_{+0}(\eta) = {}& -e\,{\ba{\cdot{\bold \eta}}\over (\eta^+)^2}\,
\de^3({\bold \eta} + \half \ba \,\eta^-) \, , \quad
{\hat \X}_{+i}(\eta) = e\,a_i {\eta^0\over (\eta^+)^2}\,
\de^3({\bold \eta} + \half \ba \,\eta^-) \, , \cr
& {\hat \X}_{0i}(\eta) = -e\,a_i{1\over \eta^+}\,
\de^3({\bold \eta} + \half \ba \,\eta^-) \, . \cr}
\eqno (5.14) $$
Since $\X_{AB}(\eta)\eta^B = 0$ (4.21) now reduces to
$\pr_B \F^{AB} = \J^A$ for this case
and, with the aid of (5.9) again, the current is determined by
$$ \eqalign{
\J^0(\eta) =  {}& e\,\Big ( 1+ {\ba{\cdot{\bold \eta}}\over \eta^+}\Big )
\de^3({\bold \eta} + \half \ba \,\eta^-) \, , \quad
\J^-(\eta) = - e\,{2\eta^0\over \eta^+} \,
\de^3({\bold \eta} + \half \ba \,\eta^-) \, , \cr
& \J^i(\eta) = e\,a_i {\eta^0\over \eta^+} \,
\de^3({\bold \eta} + \half \ba \,\eta^-) \, . \cr}
\eqno (5.15) $$
It is easy to see that this is compatible with (5.11).
\vfill\eject
\noindent{\bigbf 6 Discussion}
\medskip
It is immediately evident that the explicit elementary solution (5.1), and
also the current density given by (5.11) or (5.15), are not well defined on
$\overline {M^4}$ since they satisfy the homogeneity conditions such as (4.7)
only for $\lambda>0$ so that we can not assume $\eta^A\sim-\eta^A$. The
elementary solution for a point particle is therefore defined on the double
cover, or $S^3 \times S^1$, which has also been previously suggested as
necessary in a different context \Cast. On $S^3 \times S^1$ the trajectory
for an accelerating particle represented by ${\bold \eta}=- \half \ba \,\eta^-$
does not form a single closed curve as it would on $\overline {M^4}$. This
is demonstrated in figure 1 where the double cover corresponds to allowing
the range of $\theta,{\hat \tau}$ to be extended to $|\theta|,|{\hat \tau}| 
\le \pi$. This result is necessary since, as we made clear earlier, the
solution of Maxwell's equations given by assuming (2.13) and (2.17) are valid
on the whole of $M^4$ 
describes particles of opposite charges $\pm e$ on the two branches of the
hyperbola representing motion with constant acceleration. If a factor of
$ \ep(\Omega)$ is introduced into (2.13) and also (2.17), as would be expected
by a straightforward application of conformal transformations, then the
fields and current densities represent two particles both with charge $+e$
but there is now a discontinuity where $\Omega$ changes sign and the fields
fail to be solutions of Maxwell's equations on the light cone $\Omega =0$. This
reflects the transformation of points at infinity of the initial Coulomb
solution (2.12) corresponding to the static point charge (2.16).

Manifestly in this paper our considerations have been entirely classical but
a  natural question is whether the treatment in section 4 might be of
any assistance in discussing a conformal invariant
quantum field theory containing electrodynamics.
Making use of conformal invariance has proved an important simplification
in perturbative calculations at one and two loops \Baker\ and 
it has been the basis of alternative treatments \refs{\Adler,\Shore} where 
the Euclidean theory is quantised on $S^4$ maintaining manifest invariance
under the maximal compact subgroup $O(5)$ of the conformal group $O(5,1)$
in this case
(instead of the usual $O(4)\ltimes  T_4$ for flat space). The role of $S^4$
becomes apparent by writing the Euclidean analogue of (3.8) as
${\bold \eta}^2 + (\eta^4)^2 + (\eta^5)^2 = (\eta^6)^2 = 1 $ and, with
coordinates given by five dimensional unit vectors $(\bold \eta,\eta^4,\eta^5)$,
the covariant field strength introduced in \Adler, $\F_{ABC}, \, A=1,\dots 5$,
is defined similarly to (4.33a,b). However when the assumption of exact
conformal invariance is applied to  quantum field theories containing abelian, 
or non-abelian, gauge fields in four space-time dimensions there is no
obvious success \refs{\conf,\Fron,\confb,\confc} in deriving clear cut results. 
At a  mundane
level there are immediate problems of introducing gauge fixing while 
maintaining conformal invariance \Baker. 
More significantly attempts to impose exact conformal invariance on quantum
electrodynamics have led to the condition that current $J^\mu =0$
as an operator equation, which leads to a trivial theory. 
Presumably this reflects the standard lore of the absence of a renormalisation
group fixed point for non asymptotically free theories like quantum
electrodynamics, except when the
coupling vanishes and the fields become free. A partial explanation of
previous difficulties may be provided by the recent simple argument \Witten\ 
which shows that, in order to obtain a non trivial conformal theory with
abelian gauge fields, there must be magnetic as well as electric massless
states, although magnetic and electric charges must vanish.
Allowing for such freedom one may hope for further progress in 
understanding four dimensional conformal quantum gauge theories.
\vfill\eject
\noindent{\bigbf Appendix}
\medskip
In this appendix we describe some of the manipulations necessary to obtain
the detailed results in section 4. We initially demonstrate that $\Q^A$, defined
in (4.20), satisfies the required properties to derive the equation of
motion (4.21). It is convenient first to introduce
$$
{\hat \Q}_A(\eta) = \F_{AB}(\eta)\eta^B - \half \eta^2 \, \X_{AB}(\eta)
\eta^B \, ,
\eqno (A.1) $$
where $\eta^A$ are here assumed to be independent, without being required
to satisfy $\eta^2=0$. Using (4.11) and (4.13) we may show that
$$
\pr_A {\hat \Q}_B - \pr_B {\hat \Q}_A = 
- \half \eta^2 \, 3\pr_{[A} \X_{BC]}\eta^C \, .
\eqno (A.2) $$
This equation is then assumed to be solved, since ${\hat \Q}_A(\eta)\eta^A=0$,
by taking
$$
{\hat \Q}_A = \pr_A \Z + \quar (\eta^2)^2 \V_A \, , \qquad \V_A \eta^A = 0 \, ,
\eqno (A.3) $$
for suitable $\Z(\eta)$, homogeneous of degree zero, and $\V_A(\eta)$ which
is homogeneous of degree $-5$ and satisfies $\eta_{[A}\V_{B]}
= - {3\over 4} \pr_{[A} \X_{BC]}\eta^C + \rO(\eta^2)$. For
compatibility with (4.8) when $\eta^2=0$ we must take $\Z(\eta) = \half \eta^2
\, \Y(\eta)$ for some $\Y(\eta)$ so that (A.3) gives
$$
\Q_A(\eta) = {\hat \Q}_A(\eta) - \half \eta^2 \, \pr_A \Y(\eta) 
= \eta_A \Y(\eta) + \quar (\eta^2)^2 \V_A(\eta) \, ,
\eqno (A.4) $$
as required.

The rest of this appendix verifies the essential conditions
necessary for the consistency of (4.30). It is convenient to define, with
$\Y$ defined by (4.8) or (4.32),
$$
\I^{AB}\equiv L^{[A}{}_C \F^{B]C} + \F^{AB} - \half L^{AB} \Y \, .
\eqno (A.5) $$
The result obtained in (4.31) is then obviously
$$
\eta_B \I^{AB} = 0 \, .
\eqno (A.6) $$
For consistency of (4.30) it is also necessary that
$$
\eta^{[C}  \I^{AB]} = 0 \, .
\eqno (A.7) $$
To demonstrate that this result holds as an identity we first note that
$$
6\, \eta^{[C} L^A{}_D \F^{B]D} = - 3 L^{[A}{}_D {\tilde \F}^{BC]D} + 
{\tilde \F}^{ABC} \, , 
\eqno (A.8) $$
from the definition (4.33a), using $[\eta^C , L^{AD}] = - g^{CD} \eta^A +
g^{CA} \eta^D$ and, with the aid of (4.7), $L^A{}_D ( \eta^D \F^{BC}) = 
3\eta^A \F^{BC}$. Hence, since $\eta^{[C} L^{AB]} = 0$, it is easy to see from
(A.5) that
$$
2\, \eta^{[C}\I^{AB]} = - L^{[A}{}_D {\tilde \F}^{BC]D} + {\tilde \F}^{ABC} \, .
\eqno (A.9) $$
The vanishing of the r.h.s. of (A.9), or
$$
L^{[A}{}_D {\tilde \F}^{BC]D} = {\tilde \F}^{ABC} \, ,
\eqno (A.10) $$
is another version of the Bianchi identity which flows from the expression
(4.33b) for ${\tilde \F}^{ABC}$ since
$$ \eqalign{
g^{CD} L^{[A}{}_C L^{B]}{}_D = {}&- 2 L^{AB} \, , \cr
L^{[AB} L^{CD]} = 0 & \ \Rightarrow \ L^{[A}{}_{[C} L^{B]}{}_{D]} 
=\quar \big ( L^{AB}L_{CD} + L_{CD}L^{AB} \big) \, , \cr}
\eqno (A.11) $$
by virtue of the commutation relation (3.18) and the definition (3.17), which
leads to
$$ \eqalign {
L^{[A}{}_D L^{BC]} = {}& {\ts {1\over 3}} \big ( 2 L^{[A}{}_{\![D}
L^{B]}{}_{\!E]} - L_{DE} L^{AB} + [ L^{A}{}_{\!(D} , L^{B}{}_{\!E)}] \big )
g^{EC}  \cr
={}& - L^{[AB} \de^{C]}{}_{\! D} \, . \cr}
\eqno (A.12) $$

It remains to show how the conservation equation (4.28) also follows as a
necessary consistency condition from (4.30).\footnote{${}^{10}$}{Alternatively
from (4.21) $\eta_A \J^A = \eta_A \pr_B \F^{AB} + 2 \Y $ and using (A.1,4)
we may show that $\eta_A \J^A = {1\over 2} \eta^2 \L$ with
$\L= - \pr_A(\X^{AB}\eta_B) - \pr^2 \Y - {1\over 2} 
\eta^2 \pr_A \V^A = \pr_A \J^A + \rO(\eta^2)$, in accord with (4.27) which
then implies (4.28).} To achieve this we evaluate
$L^{A}{}_{C} \I^{BC}$ with the aid of
$$ \eqalign {
L^{A}{}_C L^B{}_D \F^{CD} = {}& - L^{[A}{}_C \F^{B]C} +  L^{(A}{}_C \F^{B)C}
- 2L^{AB} \Y + 2 g^{AB} \Y \, , \cr
L^{A}{}_C L^C{}_D \F^{BD} = {}& 3 L^{[A}{}_C \F^{B]C} + \quar L^2 \F^{AB}
- R^{(A}{}_C \F^{B)C} + 2 L^{(A}{}_C \F^{B)C} - L^{AB} \Y \, , \cr
L^{A}{}_C L^{BC} \Y = {}& R^{AB} \Y - 2 L^{AB} \Y \, , \cr}
\eqno (A.13) $$
using (4.12), (4.32), (A.11) and where
$$
R^{AB} = L^{(A}{}_C L^{B)C} \, .
\eqno (A.14) $$
From the initial definition (3.17)
$$
L^2 = - 2 (\eta{\cdot \pr})^2 - 8 \eta{\cdot \pr} \, ,
\eqno (A.15) $$
so that it is easy to see from (4.7) that
$$
L^2 \F^{AB} = 8 \F^{AB} \, ,
\eqno (A.16) $$
and hence, applying (A.13),
$$ \eqalignno{
L^{A}{}_C \I^{BC} ={}&- \I^{AB} + \half \big ( R^{(A}{}_C \F^{B)C} 
- R^{AB} \Y + L^{(A}{}_C \F^{B)C} + 2 g^{AB} \Y \big ) \cr
={}&- \I^{AB} \, , & (A.17) \cr}
$$
since the terms symmetric in $AB$ vanish. In order to demonstrate this 
it is necessary to make use of the explicit formula, from (3.17) and (A.14),
$$
R^{AB} = \eta^A \eta^B \pr^2 - (\eta^A \pr^B + \eta^B \pr^A)( \eta{\cdot \pr}
+ 1 ) - g^{AB} \eta{\cdot \pr} \, .
\eqno (A.18) $$
With this result and (4.39)
$$ \eqalignno{
R^{(A}{}_C \F^{B)C} = {}& \eta^{(A}\pr^2 \big (\eta_C \F^{B)C}\big )
- L^{(A}{}_C \F^{B)C} \cr
={}& \eta^A \eta^B \pr^2 \Y + 2\eta^{(A} \pr^{B)}\Y - L^{(A}{}_C \F^{B)C}\, , 
& (A.19a) \cr
R^{AB} \Y = {}& \eta^A \eta^B \pr^2 \Y + 2 \eta^{(A} \pr^{B)} \Y 
+ 2 g^{AB} \Y \, , & (A.19b) \cr}
$$
since, by (4.42), $\pr^2(\eta^2 \U^B) \to 0$ as $\U^B(\eta)$ is homogeneous
of degree $-3$. Subtracting (A.19b) from (A.19a) then justifies the
disappearance of the terms involving $R^{AB}$ in (A.17). 
Assuming now, in accord with (4.30) which is justified by (A.7),
$$
\I^{AB} = \eta^{[A} \J^{B]} \, ,
\eqno (A.20) $$
then, with the aid of (A.17), we obtain
$$ \eqalign {
L^{A}{}_C \I^{BC} ={}& - \half \eta^A (4+ \eta{\cdot \pr}) \J^B + \half \eta^B
L^A{}_C \J^C = - \half \big ( \eta^A  \J^B - \eta^B L^A{}_C \J^C \big ) \cr
={}& - \I^{AB} = -  \eta^{[A} \J^{B]} \, . \cr}
\eqno (A.21) $$
This immediately leads to (4.28).
\bigskip
\noindent{\bigbf Acknowledgements}
\medskip
One of us (C.C.) would like to thank Trinity College Cambridge for financial
support while this work was carried out.
\listrefs
\bye